\documentclass[%
reprint,
superscriptaddress,
 amsmath,amssymb,
 aps,
prb,
raggedbottom,
]{revtex4-2}

\usepackage{graphicx}
\usepackage{dcolumn}
\usepackage{bm}
\usepackage[T1]{fontenc}
\usepackage{placeins}
\usepackage[hidelinks]{hyperref}
\usepackage{url}
\usepackage{xcolor}
\usepackage{kotex}

\usepackage{url}

\hypersetup{colorlinks=true, urlcolor=blue, citecolor=cyan, pdfborder={0 0 0}}

\begin{document}

\title{Adiabatic confinement and number-theoretic spectra of massless Dirac fermions in twisted trilayer graphene}

\author{Kunihiro Yananose}
 \affiliation{Korea Institute for Advanced Study, Seoul 02455, Korea }
 \affiliation{Department of Physics Education, Kongju National University, Gongju 32588, Korea.}

\author{Changwon Park}
 \affiliation{ Department of Physics, Ewha Womans University, Seoul 03760, Korea }
 \affiliation{Institute for Multiscale Matter and Systems, Ewha Womans University, Seoul 03760, Korea }

\author{Young-Woo Son}
 \email{Corresponding author: hand@kias.re.kr}
 \affiliation{Korea Institute for Advanced Study, Seoul 02455, Korea }

\date{\today}

\begin{abstract}  Electrons in graphene move like massless relativistic particles, making them exceptionally mobile but difficult to trap because Klein tunnelling permits unimpeded passage through electrostatic barriers. This challenge extends to atomic collapse and hadron confinement. Twisted trilayers form Bernal and rhombohedral domains joined by smooth boundaries, bringing their carriers into contact. Here we show that a new mechanism, distinct from established approaches, confines massless fermions in equilateral triangular Bernal domains without mass barriers, bulk gaps or physical edges. Simulations and analytical modeling reveal that massless carriers have no adiabatic continuation into rhombohedral domains, while escape through nonadiabatic coupling to massive carriers is exponentially suppressed by smooth stacking variation. Perfect reflection in the adiabatic limit turns Bernal domains into quantum cavities in a gapless, edgeless metal. Striking density-of-states oscillations encode Eisenstein-integer arithmetic, linking energies and degeneracies to norms and prime factorizations. These findings bring relativistic confinement and number theory into experimentally accessible electronic spectra. \end{abstract}

\maketitle

Moir\'e materials formed from twisted bilayers of two-dimensional (2D) crystals provide versatile platforms for exploring quantum phenomena beyond those accessible in conventional solid-state systems~\cite{andrei2021NRM}. 
Extending this approach to multilayers expands the design space, enabling control over moir\'e length scales, layer number, interactions, and symmetries~\cite{Ren2025JPCM,Kennes2021NatPhys}. 
These degrees of freedom support emergent phenomena governed by physics distinct from that of the constituent layers~\cite{andrei2021NRM,Ren2025JPCM,Kennes2021NatPhys}.

Spatial variations in electronic structure provide another degree of freedom. 
Lateral interfaces between different 2D crystals enable control over band alignment, carrier injection and transport~\cite{Levendorf2012Nature,Huang2014NM,Li2015Science,Leong2018JACS}, while coupling between states with distinct dispersions and symmetries governs transmission and reflection~\cite{Nakanishi2010PRB}. 
In reconstructed moir\'e systems, smooth stacking variations create such junctions within a chemically uniform lattice.

Confining massless Dirac fermions at such interfaces is challenging because Klein tunnelling permits perfect transmission 
through smooth electrostatic barriers, obstructing conventional  confinement~\cite{Ando1998JPSJ,Katsnelson2006NatPhys}. 
Previous efforts to circumvent this limitation have explored etched structures~\cite{Ponomarenko2008Science}, confining mass terms~\cite{Berry1987PRSL,Chodos1974PRD}, and inhomogeneous magnetic fields~\cite{DeMartino2007PRL} to name a few.
Whether an intrinsic mechanism can confine these massless carriers within a metal remains an appealing question.

Twisted trilayer graphene (TTG)~\cite{zhu2020prl,Nakatsuji2023PRX,Park2025Nature} provides a natural setting for such confinement. 
Its moir\'e superstructure brings Bernal- and rhombohedral-stacked trilayer graphene (BTG and RTG) into close proximity~\cite{Park2025Nature}. 
At low energies, BTG hosts massless and massive Dirac fermions decoupled by mirror symmetry, whereas RTG exhibits a qualitatively different energy–momentum dispersion~\cite{latil2006prl,Min2008PTPS}. 
These domains invite exploration of scattering, confinement and interactions between distinct quasiparticles across atomically smooth interfaces.

\begin{figure*}[t]
\centering
\includegraphics[width=0.65\textwidth]{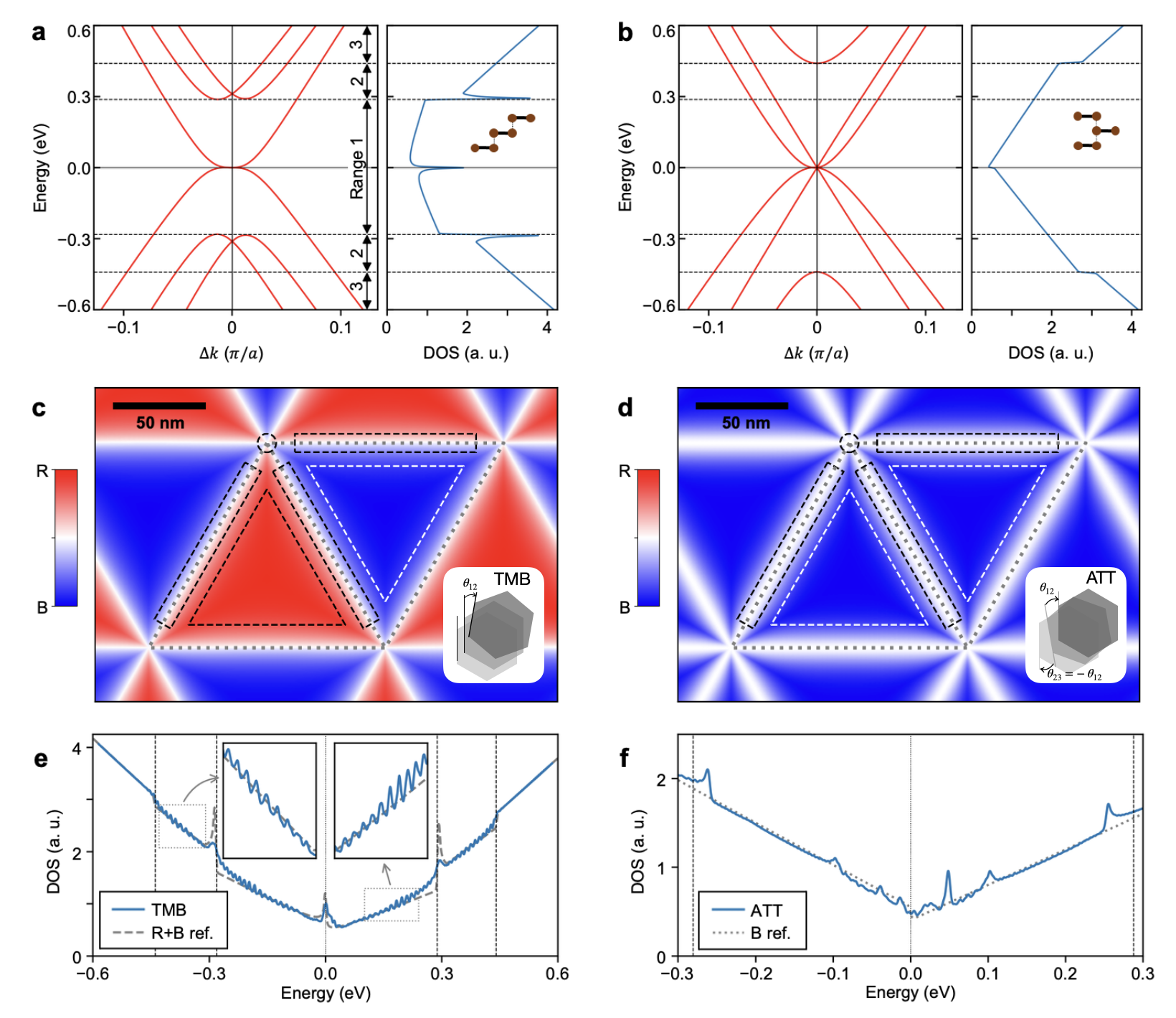}
\caption{
\textbf{Electronic structures of pristine and twisted trilayer graphene.} 
\textbf{a}, Band structures (left panel) and DOS (right panel) of RTG. The $E_F$ is set to zero. 
The energy ranges 1, 2, and 3 are marked between van Hove singularities, whose energy levels are indicated by dashed lines at $0.440,0.288,-0.281,-0.440$ eV. 
\textbf{b}, same for BTG. 
Insets in {\textbf a} and {\textbf b} show the schematic interlayer configuration.
$\Delta k$ denotes the momentum displacement from $K$, with positive (negative) values toward $M$ ($\Gamma$).
\textbf{c}, Local stacking order parameter of TMB graphene. 
Red (blue) denotes perfect rhombohedral (Bernal) stacking (see Methods). 
The dashed circle, rectangles, and triangles indicate the regions used for the projected local DOSs.
\textbf{d}, Same as \textbf{c} for ATT graphene.
Insets in {\bf c} and {\bf d} are schematic stacking configurations. 
\textbf{e} and \textbf{f}, Total DOSs of TMB and ATT shown in \textbf{c} and \textbf{d}, respectively, where the blue solid lines represent our simulations and the gray dashed or dotted lines are the reference inferred from the pristine systems.
}
\label{fig1}
\end{figure*}

We combine large-scale  simulations~\cite{Park2023NatComm,yananose2025hBN} with analytical modeling to demonstrate a new mechanism for confining massless fermions in metallic TTG, requiring neither physical edges~\cite{Ponomarenko2008Science} nor mass-gap engineering~\cite{Berry1987PRSL,Chodos1974PRD}. 
Smooth stacking boundaries break the mirror symmetry that decouples massless and massive fermions in BTG, hybridizing the two branches. 
We show that the massless branch has no propagating continuation into RTG domains and is therefore reflected adiabatically, while smooth stacking variation exponentially suppresses its nonadiabatic escape into the massive branch. 
Near-perfect backscattering thus confines massless Dirac fermions in a gapless metal. 
The computed density of states (DOS) exhibits intriguing oscillations that reveal quantization of the linear Dirac dispersion. 
The resulting spectrum encodes the arithmetic of Eisenstein integers~\cite{Pinsky1980SIAM,Itzykson1986JPA,Cox1989}, whose norms select the confined energies and whose prime factorizations determine their degeneracies. 
We believe that
TTG establishes a new route to massless-particle confinement and brings number theory into the experimentally accessible spectrum of a solid.

\subsection*{Electronic structures of twisted trilayer graphene}

Let us first discuss the stacking-dependent electronic structures of pristine systems. 
As shown in Figs.~\ref{fig1}\textbf{a} and \textbf{b},  
RTG has a pair of bands with the cubic dispersion touching at the Fermi level ($E_F$), 
whereas BTG has the decoupled linear and  quadratic dispersions~\cite{latil2006prl,Min2008PTPS}. 
Besides these low energy bands, 
two quadratic conduction and valence bands appear
beyond the van Hove singularities (vHSs) for RTG, 
and additional single quadratic band also appears beyond vHSs 
for the BTG.
Those energy levels are indicated by dashed lines in Figs.~\ref{fig1}\textbf{a} and \textbf{b}. 
The DOS 
reflects the band characteristics, such as the diverging DOS at vHSs and $E_F$ in RTG
and linear DOS with abrupt jumps at vHSs 
in BTG.

Comparing the two band structures, three distinct energy ranges separated by vHSs can be identified, 
which we denote as ranges 1, 2 and 3, respectively and mark them 
in Fig.~\ref{fig1}. 
It is noteworthy that, in ranges 1 and 2, BTG and RTG differ in the number of energy bands, with one stacking hosting an additional band relative to the other, while in range 3, the both have the same number of bands.
Owing to this mismatch, 
at least one scattering channel must be completely reflective for electrons to propagate from one stacking to the other when they are joined side by side (See Method for details)~\cite{Glazman1988JETP}. 
Such a situation arises naturally in twisted trilayer systems.

\begin{figure}[t]
\centering
\includegraphics[width=0.43\textwidth]{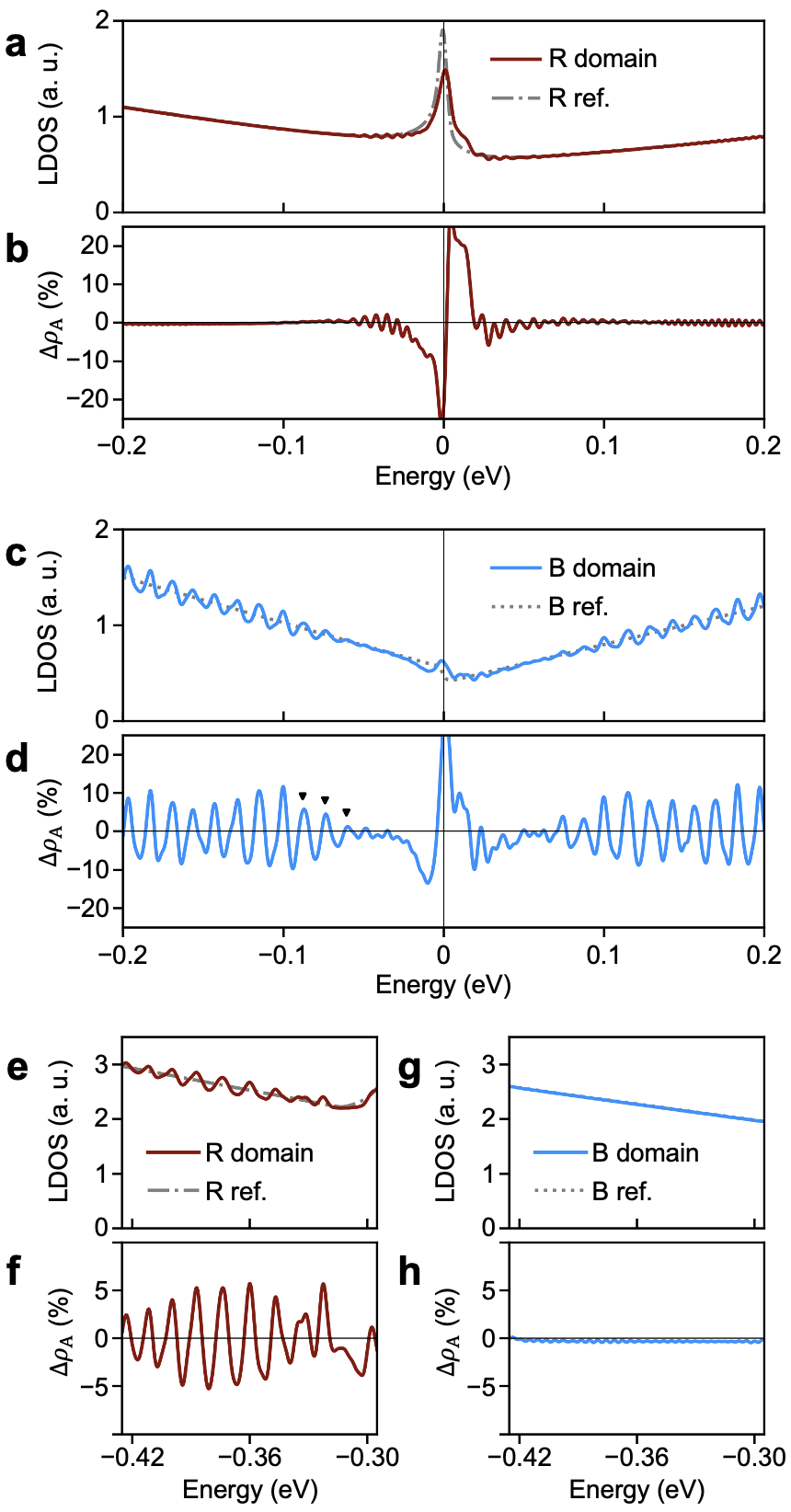}
\caption{
\textbf{Dichotomy of local density of states in triangular domains.}
\textbf{a}, The LDOS ($\rho_A(E)$) on the triangular region (Fig.~\ref{fig1}\textbf{c}) of the rhombohedral domain in energy range 1 is shown as a solid line. 
For comparison, the reference DOS of the RTG is plotted by a gray dash-dot line. 
\textbf{b}, The normalized deviation of LDOS in {\bf a} from the reference DOS, $\Delta\rho_A(E)$ (defined in the text).
\textbf{c} and \textbf{d}, The same quantities as in {\textbf a} and {\textbf b}, but for the Bernal domain in Fig.~\ref{fig1}\textbf{c}. 
\textbf{e} and \textbf{f}, The same data as in \textbf{a} and \textbf{b}, but over energy range 2 for the rhombohedral domain.  
\textbf{g} and \textbf{h}, Analogously, these panels show the Bernal domain LDOS over the range 2.
}
\label{fig2}
\end{figure}

To investigate the scattering behavior, we consider two representative TTGs characterized by twist angles $\theta_{12}$ and $\theta_{23}$ between adjacent layers.
The first is twisted mono-bilayer (TMB) graphene, formed by twisting a monolayer relative to a bilayer by $\theta_{12}$. Its fully relaxed structure consists of alternating RTG and BTG domains separated exclusively by RTG--BTG boundaries (Fig.~\ref{fig1}\textbf{c}). 
For $\theta_{12}<0.05^\circ$, these triangular domains undergo spontaneous symmetry breaking into a kagome-like pattern~\cite{Park2025Nature}. 
The second is alternately twisted trilayer (ATT) graphene with $\theta_{23}=-\theta_{12}$. 
After relaxation, it forms alternating triangular Bernal domains (ABA and ACA), as shown in Fig.~\ref{fig1}\textbf{d}.
We note that, for angles around 1--5 degrees, renormalization of band structures can occur owing to moir\'e potentials~\cite{zhu2020prl, Nakatsuji2023PRX}. 
So, we adopt a marginal twist angle of $\theta_{12}=0.11^\circ$ for both systems (corresponding to a moir\'e period of $L_M=1280.13~\text{\AA}$), thereby minimizing the strong band renormalization~\cite{Park2025Nature}
(for detailed methods, See Methods).

Figure~\ref{fig1}\textbf{e} shows the DOS of the TMB system obtained from large-scale atomistic calculations, together with a reference constructed from the averaged DOSs of pristine RTG and BTG. 
Overall, the calculated DOS closely follows the reference, reproducing the singularity at $E_F$ and other vHSs. 
Notably, however, ranges 1 and 2 develop oscillatory features, 
whose envelope approximately follows the reference DOS. 
By contrast, range 3 exhibits excellent agreement with the reference and no discernible oscillations.
The DOS of ATT graphene is qualitatively different. 
As shown in Fig.~\ref{fig1}\textbf{f}, the DOS agrees well with that of pristine Bernal stacked trilayer over a broad energy range, yet additional peaks emerge at a few energies.

We have verified that the oscillations in Fig.~\ref{fig1}{\bf e} are not artifacts of incomplete convergence (see Methods). They therefore indicate that the electronic structure of TMB differs qualitatively from those of pristine trilayer and ATT systems. 
This is unexpected because the twist angle is small enough to avoid moir\'e-driven band renormalization, suggesting another origin.

\subsection*{Oscillatory density of states in triangular domains}


To elucidate origin of the intriguing oscillations, we examine the local density of states (LDOS) of $\rho_A(E)$, projected onto specific regions $A$, indicated by dashed lines in Fig.~\ref{fig1}\textbf{c}. 
In Fig.~\ref{fig2}, we present the enlarged views of the $\rho_A(E)$, together with its normalized value defined as $\Delta\rho_A(E) = (\rho_A(E)-\rho_{\text{R}}(E))/\rho_{\text{R}}(E)$ for various local domains of the TMB graphene across energy ranges 1 and 2. 
Here, $\rho_{\text{R}}(E)$ denotes the reference DOS of pristine BTG (RTG) for the blue (red) triangular domain in Fig.~\ref{fig1}{\bf c}. 

The oscillatory behavior of $\Delta\rho_A(E)$ reveals a clear dichotomy between the two domains.
In range 1,  
the Bernal domain shows pronounced  oscillations that persist throughout the energy range (Figs.~\ref{fig2}{\bf c} and {\bf d})
while rhombohedral domain does not (Figs.~\ref{fig2}{\bf a} and {\bf b}). 
Beyond the first vHS, the behavior reverses. 
In range 2, the Bernal domain becomes featureless (Figs.~\ref{fig2}{\bf g} and {\bf h}), while the rhombohedral domain develops similarly strong oscillations (Figs.~\ref{fig2}{\bf e} and {\bf f}). 
We therefore identify the DOS oscillations in Fig.~\ref{fig1}{\textbf e} as originating from the Bernal domain in range 1 and from the rhombohedral domain in range 2. 
By contrast, no discernible oscillatory features appear in range 3 (Fig.~\ref{fig1}{\bf e}).
As discussed, these contrasting behaviors are closely linked to the different numbers of available bands in each domain 
at a given energy.

As mentioned already, two connected conductors with different numbers of bands have perfect reflection channels at their interface~\cite{Glazman1988JETP}. 
In range 1, a Bernal domain has one more conducting channel than a rhombohedral domain, so one BTG channel should undergo perfect reflection (See Methods). 
In the layer-resolved LDOS shown in Extended Data Fig.~\ref{EFig_layerLDOS}, the middle layer shows no oscillations, whereas the top and bottom layers do. 
This strongly suggests that the confined state is the massless odd mode in ABA stacking, since this mode resides only on the top and bottom layers~\cite{Min2008PTPS}. 
As will be demonstrated below, this is indeed the case. 
In range 2, confinement shifts to the RTG. Here, the perfectly reflected channel arises from hybridized bands rather than a single band (not shown here). 
In the following, we focus on range 1, where massless-particle confinement occurs.

The LDOS in Fig.~\ref{fig2}\textbf{d} exhibits four notable features.
First, the peak spacing in range 1 is remarkably uniform, with an average value of $13.60 \pm 0.91$ meV.
Second, the noticeable oscillations in quantized spectra emerge only away, about $\pm 60$ meV, from $E_F$ as shown in Fig.~\ref{fig2}\textbf{c}, not starting from $E_F$. 
Third, beyond the primary oscillation, the peak amplitudes exhibit a slower beating-like envelope modulation. 
Fourth, the DOS between peaks remains only weakly suppressed, in sharp contrast to conventional quantum dots 
~\cite{Banin1999Nature}. 
As will be shown, these features are  hallmarks of quantization of massless fermions confined in a triangular domain of TTG.

In the ATT system, the LDOS within the Bernal domains  exhibits no discernible oscillations (Extended Data Fig.~\ref{EFig_LDOS_ATT}\textbf{a} and {\bf b}). 
By contrast, $\Delta\rho_A(E)$ projected onto the domain boundaries displays several pronounced peaks (Extended Data Fig.~\ref{EFig_LDOS_ATT}\textbf{c} and {\bf d}), indicating that the deviations observed in Fig.~\ref{fig1}{\textbf f} originate primarily from the domain boundaries rather than the domain interiors. 
For both TMB and ATT systems, $\Delta\rho_A(E)$ projected onto the vertex regions exhibits several strong peaks near $E_F$, accounting for features that deviate from the otherwise boundary-state signatures and regular oscillatory behavior (Extended Data Fig.~\ref{EFig_LDOS_ATT}\textbf{e} and Extended Data Fig.~\ref{EFig_LDOS_TMB}\textbf{c}).

\begin{figure}[t]
\centering
\includegraphics[width=0.45\textwidth]{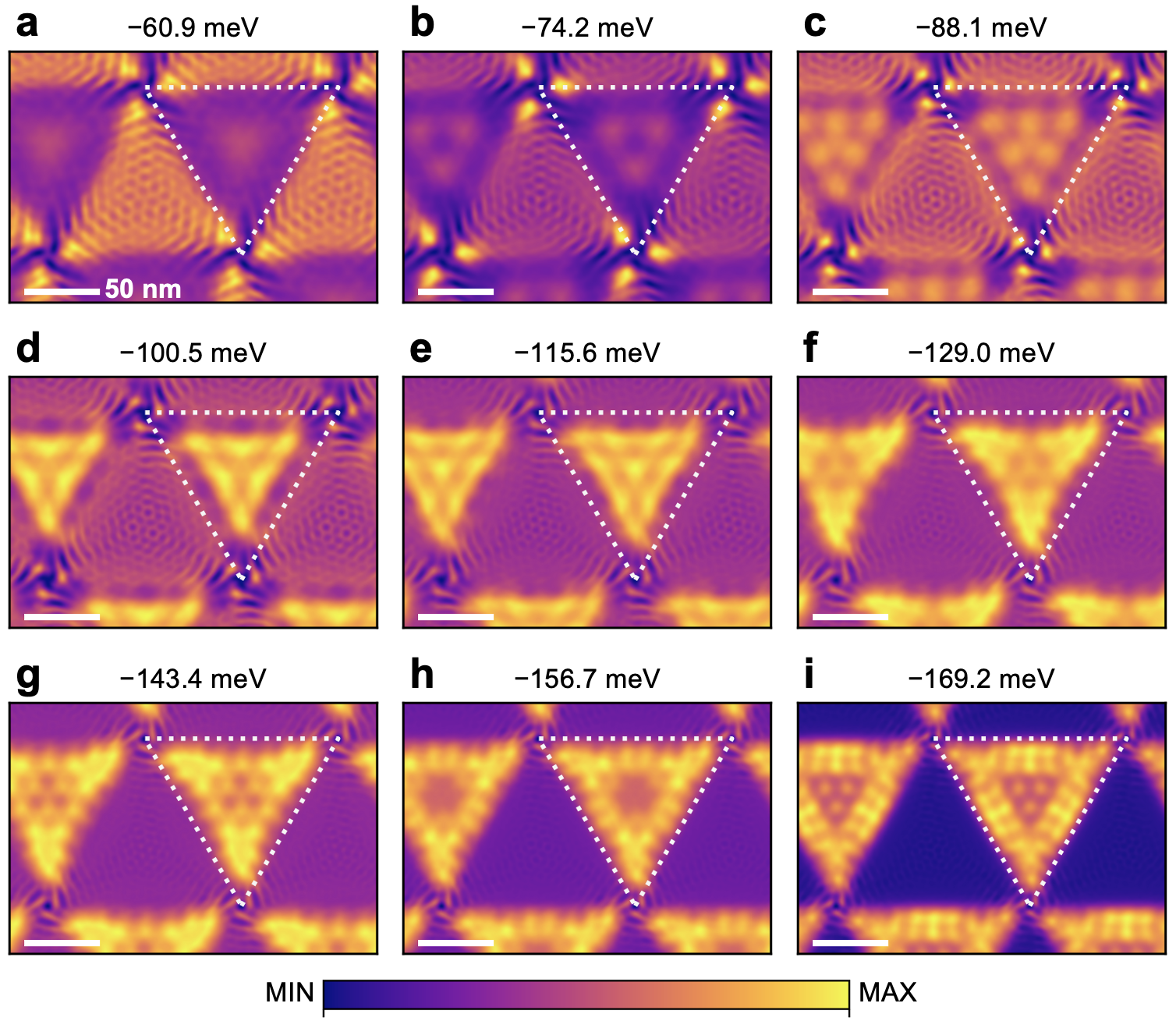}
\caption{
\textbf{Energy- and layer-resolved electron density distributions of the TMB system.} 
The energy level $E$ above each panel is where the density $\rho_{E}(\mathbf{r})$ is evaluated for peaks of oscillatory DOS in Fig.~\ref{fig2}{\bf d}, where the first three peaks corresponding to \textbf{a}-\textbf{c} are indicated by markers. 
$\rho_{E}(\mathbf{r})$ represents the sum of the layer resolved densities of layer 1 and layer 3.
The dotted line triangle indicates the Bernal domain.
The colormap range is individually scaled to the minimum and maximum density values of each panel. Note that the minimum is a finite value as the system is semimetallic.
Each panel's a ratio of MIN/MAX value is: 
\textbf{a} 0.505, \textbf{b} 0.521, \textbf{c} 0.645, 
\textbf{d} 0.645, \textbf{e} 0.593, \textbf{f} 0.570, \textbf{g} 0.555, \textbf{h} 0.559, and \textbf{i} 0.601.
}
\label{fig3}
\end{figure}

\subsection*{Charge densities of confined triangular states}

To visualize the confined states, we calculate the energy- and layer-resolved charge density $\rho_E(\mathbf r)$ for several  peaks of the oscillatory DOSs (see Methods). 
Figures~\ref{fig3} show $\rho_E(\mathbf r)$ for the first nine consecutive peaks in Fig.~\ref{fig2}\textbf{d}. 
We display a sum of top and bottom layers for each peak since the massless states are only on these layers not on the middle one.  
The detailed $\rho_E(\mathbf r)$ for each layer and corresponding scanning tunneling microscopy image simulations can be found in Extended Data Fig.~\ref{EFig_rho_layer}.

As shown in Fig.~\ref{fig3} and Extended Data Fig.~\ref{EFig_rho_layer}, the Bernal domain exhibits pronounced nodal patterns, with the number of nodes increasing with energy, providing direct evidence of quantum confinement. 
Notably, the effective area occupied by the confined charge density is smaller than the geometric boundary between the two stacking domains (indicated by the white dotted lines) and expands as the state energy moves away from $E_F$. 
At the same time, boundary charge accumulation is observed along the geometric boundaries within the middle layer of the Bernal domain, as shown in Extended Data Figs.~\ref{EFig_rho_layer}\textbf{g}--\ref{EFig_rho_layer}\textbf{l}, of which origin will be discussed later. 

Unlike the well-defined confined states in Bernal domain, the rhombohedral domain displays more uniform wave patterns, consistent with hybridized states with neighboring domains rather than confined states. 
We further find that the largest amplitudes for first few low-energy states are concentrated near the vertices, and gradually diminish as the energy moves away the $E_F$.

Similarly, Extended Data Fig.~\ref{EFig_ran2} shows $\rho_E({\bf r})$
corresponding to DOS peaks in range $2$ of Fig.~\ref{fig2}\textbf{f}. 
These states also clearly exhibit a strong confinement effect, but now confined only within the rhombohedral domain.
We find that all the other DOS peaks also closely follow the nodal structures expected from confinement inside triangular domain as well as the confinement rules dictated by excess number of bands discussed above.  
In addition, we identify the boundary and vertex states contributing a couple of LDOS peaks of TMB and ATT systems 
as shown in Extended Data Figs.~\ref{EFig_rho_TMB} and~\ref{EFig_rho_ATT}, respectively.

\subsection*{Boundary as adiabatic confinement barrier}

To analyze the quantized states in Bernal domains, we consider massless Dirac fermions confined to an equilateral triangle. 
The nonrelativistic counterpart has a long history, dating back to Lam\'e~\cite{Lame1852} and later approaches~\cite{Mathews1970,Pinsky1980SIAM,Krishnamurthy1982JPA,Itzykson1986JPA}. The confined states found here are massless, however, so Schr\"odinger solutions with Dirichlet boundaries do not apply. 
Our simulations reveal out-of-plane boundary corrugations below $0.02$~\AA~and in-plane profiles extending about $124$~\AA~(Extended Data Fig.~\ref{EFig_height}), ruling out valley mixing~\cite{mccann2004jphyscond,Brey2006PRB}. 
Moreover, no infinite mass exists outside the domain, unlike the Berry--Mondragon (BM) condition~\cite{Berry1987PRSL,Dietz2021APP} for neutrino billiards or the MIT bag model~\cite{Chodos1974PRD} for hadron confinement.

The confinement found in our  simulations follows a different route from established ways of trapping a massless Dirac 
particle 
\cite{Ponomarenko2008Science,mccann2004jphyscond,Brey2006PRB,Berry1987PRSL}. 
We find that the boundary here does not add any mass to the odd ABA state (See Methods for proof). 
Instead, 
it turns out that the absence of an adiabatic continuation is origin of massless particle confinement. 
On the ABA side there are two propagating channels, but on the ABC side only one.
We explicitly show that this single exit carries the massive even ABA mode adiabatically across the smooth wall (See details in Methods). 
By unitarity, the massless mode has nowhere to go and turns back. Reflection therefore follows from channel counting and smoothness, becoming exact at every incidence angle in the adiabatic limit.

Two unusual features follow. 
First, confinement improves with wall smoothness,
while a sharp interface would leak appreciably (Methods). 
The lattice relaxation that creates the domains is therefore also what makes them quantum cavities.
Second, the trapped fermions retain their linear dispersion, so the cavity is a true massless billiard with a current-free boundary (Methods). Although the microscopic mechanism is entirely different, the BM condition~\cite{Berry1987PRSL} emerges here as a phenomenological description of our adiabatic confinement.

\begin{figure*}[t]
\centering
\includegraphics[width=0.700\textwidth]{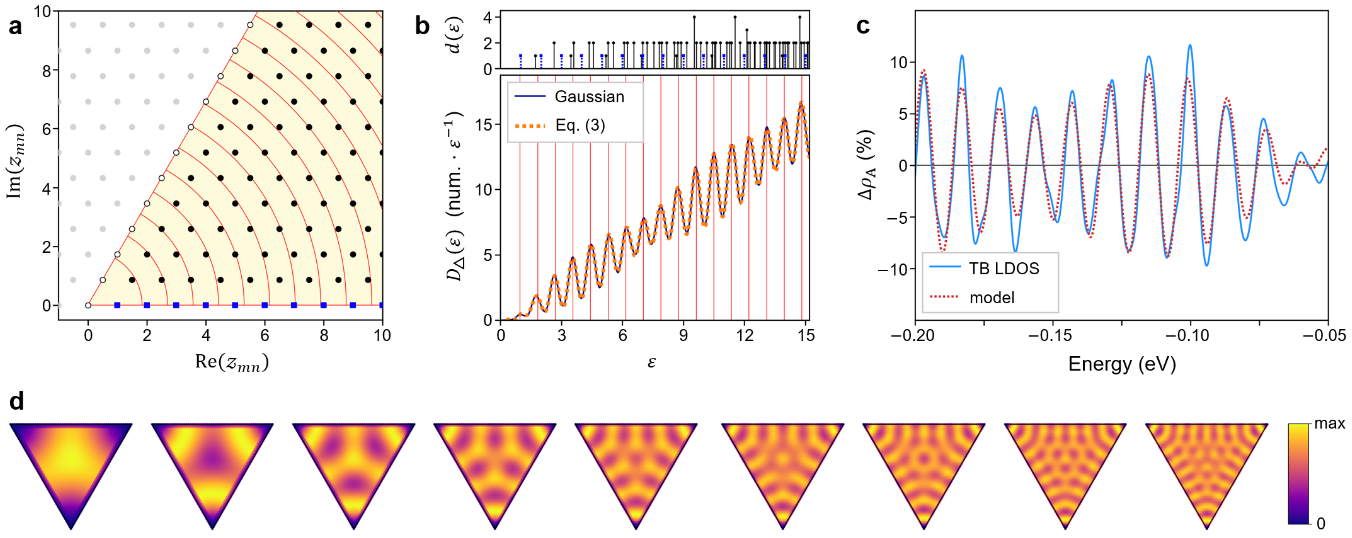}
\caption{
\textbf{Bound massless states in a triangular cavity} 
{\textbf a}, Eisenstein numbers, $z_{mn}=m+n\omega$, satisfying $m> n\ge 0$, 
are marked by filled circles in shaded region of the complex plane.  
The $z_{mn}$ on the excluded ($m=n$) and included ($n=0$) boundaries are denoted by the empty circles and the blue squares, respectively.
{\textbf b}, 
$d(\varepsilon)$ is a number of degeneracy for a given energy $\varepsilon$ (upper panel)
and DOS (low panel) as a function of $\varepsilon$, $D_\triangle(\varepsilon)$.
The degeneracy $d(\varepsilon)$ originating from the filled circles (blue squares) in {\textbf a} are shown by black solid (blue dotted) lines.
The solid line is computed by a Gaussian convolution ($\sigma=0.2$) of $d(\varepsilon)$ and the dotted line is from Eq.~\eqref{Eq:ReDOS} with the same $\sigma$.
In {\textbf a} and {\textbf b}, the red solid lines represent the quasiperiodic levels $\varepsilon= M\sqrt{3}/2+\sqrt{3}/16$ ($M= 1,2,3,\cdots$).
{\textbf c} The normalized deviation of LDOS of BTG domain from simulations (solid line) in Fig.~\ref{fig2} {\textbf d} and from our model (dotted line). 
{\textbf d}, $\rho_\triangle({\bf r};\varepsilon)$ of the first nine DOS peaks shown in \textbf{b} arranged in ascending energy order from the left to right.  
The densities are integrated using a Gaussian weight ($\sigma=0.2$) centered at the energy of the peaks.
}
\label{fig4}
\end{figure*}

\subsection*{Ideal Dirac Billiard and Eisenstein Spectrum}

To reveal the spectral consequences of this confinement mechanism, we first consider massless Dirac fermions in an equilateral triangle of side $L_M$ subject to the BM boundary condition. 
The resulting ideal spectrum provides a natural reference for understanding how confinement is reshaped within the actual moir\'e domains.

The eigenproblem is $\mathcal H_\triangle\Psi=\varepsilon\Psi$, with
$\mathcal H_\triangle=-i\hbar v_F(\sigma_x\partial_x+\sigma_y\partial_y)$,
where $\Psi$ is a two-component sublattice spinor, $v_F$ is the Fermi velocity, and $\sigma_{x,y}$ are Pauli matrices acting in sublattice space. 
Smooth boundaries allow us to consider a single valley. 
The positive-energy eigenstates $\Psi_{mn}$ have energies
$
\varepsilon_{mn}=\varepsilon_0|z_{mn}|
$
where $\varepsilon_0=\frac{4\pi\hbar v_F}{3L_M}\simeq18.82$ meV, $z_{mn}=m+n\omega$ is an Eisenstein integer, $\omega=e^{2\pi i/3}$, and $|z_{mn}|^2=m^2-mn+n^2$~\cite{gaddah2018JphysA, Dietz2021APP}. 
The BM boundary condition selects the integer pairs
$\mathcal D=\{(m,n)\mid m>n\ge0\}$,
which occupy one-sixth of the triangular lattice of Eisenstein integers, as shown in Fig.~\ref{fig4}{\bf a} (see Methods).

This correspondence places arithmetic structure of Eisenstein number directly within the confinement spectrum~\cite{Pinsky1980SIAM,Itzykson1986JPA,Cox1989}. 
Using the dimensionless energy $\varepsilon=E/\varepsilon_0$, the discrete DOS is
$d(\varepsilon)=\sum_{(m,n)\in\mathcal D}\delta(\varepsilon-|z_{mn}|)$.
Although its level multiplicities appear irregular in the upper panel of Fig.~\ref{fig4}{\bf b}, Gaussian broadening reveals a clear oscillatory pattern. 
To understand its origin, we introduce the counting function $n_\triangle(\varepsilon)$, the number of allowed Eisenstein integers satisfying $|z_{mn}|\le\varepsilon$. 
The relation
$d(\varepsilon)=\partial_\varepsilon n_\triangle(\varepsilon)$
then connects the DOS to the distribution of lattice points within a circle. 
Using the prime-factorization properties of $z_{mn}$~\cite{Pinsky1980SIAM,Cox1989}, 
we obtain the following approximation for the Gaussian-broadened DOS,
\begin{equation}
d_\triangle(\varepsilon)\simeq
\frac{4\pi}{\sqrt{3}}\varepsilon
\left[
\frac{1}{6}
+e^{-8\pi^2\sigma^2/3}
J_0\left(\frac{4\pi}{\sqrt{3}}\varepsilon\right)
\right],
\label{eq:DOS}
\end{equation}
where $\sigma$ is the dimensionless Gaussian width and $J_0$ is the zeroth-order Bessel function of the first kind (see Methods). 
The linear background reflects the massless dispersion, while the Bessel term produces confinement oscillations with an asymptotic energy spacing of $\sqrt{3}\varepsilon_0/2$. 
These features capture the principal structure of the simulated DOS in Fig.~\ref{fig2}{\bf c}, linking its oscillations to the arithmetic of the triangular confinement spectrum.

\subsection*{Self-energy corrections from leakages near $E_F$}

The actual moir\'e domains, however, depart from isolated triangular cavities~\cite{gaddah2018JphysA,Dietz2021APP}
owing to two energy dependent leakage channels.
First, one arises from ABB-stacked regions of area $\sim l^2$ at the corners of the domains, that  depends on the overlap between ABB states and the massless mode of ABA stacking. 
Below $E_c=44.7$~meV, the relevant ABB bands have strong odd character, with $\langle P_o\rangle=$0.8--0.9, where $P_o$ projects onto the ABA mirror-odd subspace (Extended Data Figs.~\ref{Efig_ABB}{\bf a} and {\bf b}; see Methods). 
Their momenta are also comparable to those of the incident massless mode, providing a matching channel through which carriers can escape. We therefore expect corner leakage to obstruct confinement below $E_c$. 
At this energy, the matching ABB band reaches its maximum and the channel closes. Above $E_c$, the remaining states lie on an outer branch with momenta several times larger than those of the incident mode and $\max\langle P_o\rangle\simeq0.05$. 
This mismatching 
suggests a reduction in corner transmission by more than an order of magnitude, allowing well-defined cavity resonances to emerge.

The other involves modes propagating parallel to the triangle sides, which can couple to massive states without undergoing reflection from the boundaries.
This residual leakage channel remains for the $n=0$ modes, marked by blue rectangles in Fig.~\ref{fig4}{\bf a}. 
These modes propagate parallel to the triangle sides without encountering the wall. 
Nevertheless, the varied interlayer coupling within the boundary mixes them with middle-layer states and thereby connects them to the massive branch. 
Only about $26\%$ of their weight remains within the domain (see Methods). Their coupling to massive states is also visible in the enhanced line-shaped intensity along the boundaries in the layer 2 of $\rho_E({\bf r})$ (Extended Data Figs.~\ref{EFig_rho_layer}{\bf g}--{\bf l}).

To incorporate these departures from ideal confinement, we adopt a minimal affine self-energy that shifts and rescales the bare spectrum,
\begin{equation}
E_{mn}\simeq E_c+
\frac{4\pi}{3}
\frac{\hbar v_{\mathrm{eff}}}{L_M}
\sqrt{m^2-mn+n^2},
\label{eq:Triangle_Energy}
\end{equation}
where $v_{\mathrm{eff}}=Zv_F$ and $Z(<1)$ is the renormalized factor due to low energy leakage sources.  
Here, $E_c$ serves as the spectral offset associated with the suppression of corner leakage (see Methods). 
With the renormalized dimensionless energy
$\tilde\varepsilon=\tfrac{E-E_c}{Z\varepsilon_0}$
and the retained weight $A_w=0.26$ of the $n=0$ modes, 
we obtain the renormalized DOS,
\begin{equation}
D_\triangle(\tilde\varepsilon)\simeq
Z d_\triangle(\tilde\varepsilon)
+Z(A_w-1)b_{\mathrm{wall}}(\tilde\varepsilon),
\label{Eq:ReDOS}
\end{equation}
where
$
b_{\mathrm{wall}}(\tilde\varepsilon)\simeq
1+2e^{-2\pi^2\sigma^2}\cos(2\pi\tilde\varepsilon)
$
is the DOS of the boundary-parallel modes. 
The second term reduces their contribution to account for leakage (See Methods for detailed derivations), 
such that Eq.~\eqref{Eq:ReDOS} closely reproduces the simulated DOS in Fig.~\ref{fig4}{\bf b} and yields the normalized oscillatory contribution $\Delta\rho_A(E)$. 
The fast oscillation has a periodicity of
$\tfrac{\sqrt3}{2}Z\varepsilon_0\simeq13.6$~meV,
while its superposition with the $n=0$ modes produces a slower beating pattern. 
Both features agree quite well with our simulations, as shown in Figs.~\ref{fig4}{\bf b} and {\bf c}. 
Leakage thus modifies the weights and energy scale of the resonances while preserving the arithmetic structure underlying the DOS.

The same description also accounts for the spatial signatures of confinement. 
Each eigenstate gives a probability density
$\rho_{mn}(x,y)=|\Psi_{mn}(x,y)|^2$
(see Methods for explicit forms). 
Summing these densities over the allowed integer pairs, with Gaussian broadening centered on the corresponding energies $E_{mn}$, yields the energy-resolved distribution $\rho_\triangle({\bf r};E)$ (see Methods). The resulting patterns in Fig.~\ref{fig4}{\bf c} closely reproduce those obtained from the atomistic simulations in Fig.~\ref{fig3}, connecting the arithmetic organization of the spectrum to the real-space structure of the confined Dirac states.

\subsection*{Discussions}

Taken together, our million-atom simulations and analytical modelling establish that massless Dirac fermions can be confined within metallic domains without surrounding semiconducting barriers. 
Smooth stacking variations act as internal reflectors to massless carriers when a mirror-protected mode loses its propagating continuation and conversion into other channels is suppressed. 
The diverse energy–momentum dispersions of multilayer graphene~\cite{Min2008PTPS,Koshino2013PRB2} suggest that this mechanism may extend to other multilayer moir\'e systems with suitable channel structures.
Twisted trilayers already host a rich variety of stacking domains~\cite{Ren2025JPCM,Park2025Nature}, offering a natural setting in which geometry and quasiparticle character can be brought together to design confinement spectra.

Considering that manipulations of local and global boundary shapes~\cite{Yankowitz2014NatMater,Jiang2018NatNano} 
are already demonstrated, 
this engineering possibility raises the prospect of reshaping quantum cavities within a continuous crystal. 
Deforming triangular domains could reveal how the degeneracies organized by Eisenstein integers evolve with geometry, while suitable boundary shapes could open a window onto relativistic quantum chaos~\cite{Berry1987PRSL,Huang2018PR}. 
Spatially resolved tunnelling spectroscopy could follow this evolution through discrete resonances and their local DOS patterns. 
Tracking the level spacings as domain size varies would further test the geometric origin of quantization, bringing the relation between shape, arithmetic and quantum dynamics within experimental reach.

\section*{Methods}\label{sec:methods}

\subsection*{Computational methods}

The fully relaxed atomic structures were calculated using the interatomic potential method developed for efficient multi-million atom simulations~\cite{Park2023NatComm}.
For intralayer interactions, the harmonic potential was fitted against the phonon dispersion obtained from the first principles calculations~\cite{Park2025Nature}.
For interlayer interactions, we adopted the Kolmogorov-Crespi potential~\cite{Kolmogorov2005PRB} combined with long-range inter-atomic forces across all three layers, which is necessary to reproduce the energetic relation between the different stacking types~\cite{Park2025Nature}.

Electronic structures were calculated using the large-scale tight-binding (TB)  method used in our previous work~\cite{Park2025Nature, yananose2025hBN}. For the TB Hamiltonian model, we followed the previous study~\cite{Park2025Nature}, where the Slater-Koster hopping parameter between the $p_z$ orbitals of two carbon atoms displaced by $\mathbf{r}$ is given as $t(\mathbf{r}) = n^2V_{\sigma}(r) + (1-n^2)V_{\pi}(r)$, where $n = z/r$ ($r=\left|\mathbf{r}\right|$). The decaying hopping strengths are $V_{\sigma(\pi)}(r) = \gamma_{\sigma(\pi)} \exp\left[q_{\sigma(\pi)}\left(1-r/a_{\sigma(\pi)}\right)\right]$, where $\gamma_{\sigma} = 0.48$ eV, $\gamma_{\pi} = -2.7$ eV, $a_{\sigma} = 3.349$ \AA, $a_{\pi} = 1.418$ \AA, and $q_{\pi}/a_{\pi} = q_{\sigma}/a_{\sigma} = 2.218$ \AA$^{-1}$~\cite{Park2025Nature}. 
The nearest neighbors are considered for the intralayer hoppings, whereas the hoppings within the cut-off distance 5 \AA\ are considered for the interlayer interactions (the nearest interlayer hopping is given by $\gamma_1=390$ meV). 
We find that increasing cut-offs only changes the very low energy bands near the Fermi level ($E_F$), not altering any signature above.
The ARPACK iterative solver~\cite{lehoucq_arpack_1998, opencollab_arpack-ng} is adopted to obtain the eigen-solutions of the multi-million-dimensional TB Hamiltonian matrix. We examined only certain energy ranges near $E_F$ instead of the full energy spectrum. This is possible by performing the shift-invert mode calculation of the Lanczos scheme.

To obtain a high-resolution density of states (DOS), we calculated eigenstates on a $6\times6$ $\Gamma$-centered $k$-space grid for twisted trialyer graphene (TTG) with a moir\'e unit-cell length of 1280.13~\AA. 
The DOS was obtained by applying Gaussian broadening to the eigenenergies with a broadening factor of $\sigma = 2$ meV. 
By varying both the $k$ grid and $\sigma$, we carefully tested the convergence of the DOS to exclude spurious wiggles arising from insufficient eigenenergy sampling, while also avoiding artificial smoothing of genuine DOS variations.
We find that the domain-resolved local DOS is well converged to the untwisted reference at $\sigma = 2$ meV, while keeping its physical fine structures. In contrast,  the larger broadening of 6 meV 
can suppress the physical oscillations~\cite{Park2025Nature}.

The Gaussian convoluted energy-resolved spatial density around $E$
is defined as 
$\rho_{E}(\mathbf{r}) 
=\sum_{n}\int_\text{BZ} 
w_\sigma(E-E_{n\mathbf{k}})
|\psi_{E_{n\mathbf{k}}}(\mathbf{r})|^2  d^2 k$ 
where 
$w_\sigma(E)=\frac{1}{\sqrt{2\pi}\sigma}e^{-
E^2/
{2\sigma^2}}$
and  
$\sigma=3$ meV is used
unless specified elsewhere.

\subsection*{Stacking order parameter for trilayer graphene}

We quantify the stacking orders for arbitrary trilayer stackings by approximating the local  structure as a relative sliding of untwisted layers. 
Following a previous work for twisted bilayer graphene~\cite{alden2013pnas},
we first introduce four parameters: $u^{2\text{A}}_{1\text{B}}$, $u^{2\text{B}}_{1\text{A}}$, $u^{3\text{A}}_{2\text{B}}$, and $u^{3\text{B}}_{2\text{A}}$, where numbers denote layers (1: bottom, 2: middle, 3: top), and A/B are sublattices. 
Each represents the shortest in-plane distance from the atom labeled by the superscript to the atom by the subscript 
(e.g., $u^{2\text{A}}_{1\text{B}}$ is  the shortest in-plane distance from middle layer atom on A sublattice to bottom layer one on  B sublattice). 
They are evaluated at each atom and interpolated on a regular grid. 
For unstrained unit-cell parameters, the maximum value should be the bond length of the hexagonal lattice, $u_{\text{max}} = a_0$. 
The local trilayer stacking order are then,
\begin{eqnarray}\label{eq:stackorderparam}
u_{\text{ABC}} &=& (1-u^{3\text{A}}_{2\text{B}}/u_{\text{max}})(1-u^{2\text{A}}_{1\text{B}}/u_{\text{max}}) \nonumber \\
u_{\text{ACB}} &=& (1-u^{3\text{B}}_{2\text{A}}/u_{\text{max}})(1-u^{2\text{B}}_{1\text{A}}/u_{\text{max}}) \nonumber\\
u_{\text{ABA}} &=& (1-u^{3\text{B}}_{2\text{A}}/u_{\text{max}})(1-u^{2\text{A}}_{1\text{B}}/u_{\text{max}}) \nonumber\\
u_{\text{ACA}} &=& (1-u^{3\text{A}}_{2\text{B}}/u_{\text{max}})(1-u^{2\text{B}}_{1\text{A}}/u_{\text{max}})\nonumber. 
\end{eqnarray}
Each equals 1 in its corresponding stacking and 0 when another parameter reaches 1. 
To distinguish rhombohedral-stacked trilayer graphene (RTG) from Bernal trilayer graphene (BTG), we define
\begin{equation}
u_{\text{RB}} = u_{\text{ABC}} + u_{\text{ACB}} - u_{\text{ABA}} - u_{\text{ACA}},\nonumber
\end{equation}
which equals $+1(-1)$ for RTG (BTG). Here each subscript denotes a specific stacking sequence of trilayer.

\subsection*{Existence of perfectly reflected channels}

Let us suppose that 
BTG and RTG occupy the regions $x<-l$ and $x>l$, respectively, with a smooth boundary in $|x|\le l$ that is uniform along $y$. Then $k_y$ is conserved, reducing scattering at fixed $E$ and $k_y$ to a one-dimensional problem.
From Extended Data Figs.~\ref{EFig_height}{\bf b} and {\bf d}, the full
width at half maximum of the stacking gradient is about 124~\AA, so $l\simeq62$~\AA.
In the range 1 of Fig.~\ref{fig1}{\bf a}, the BTG on the left has two propagating channels
($n_L=2$) while the RTG on the right has one ($n_R=1$).
One can construct a scattering matrix $S$, defined by $\psi_\text{out}=S\psi_\text{in}$,
that connects the states incoming to the boundary, $\psi_\text{in}$, to the states outgoing
from it, $\psi_\text{out}$~\cite{Fisher1981PRB}. 
In matrix form,
\begin{equation}
S=
\begin{pmatrix}
r & t' \\
t & r'
\end{pmatrix}. \nonumber
\end{equation}
Here,  $\psi_\text{in}=(\phi^L_1, \cdots,\phi^L_{n_L},\phi^R_1,\cdots,\phi^R_{n_R})^{\top}$ and
$\psi_\text{out}=(\varphi^L_1,\cdots,\varphi^L_{n_L},\varphi^R_1,\cdots,\varphi^R_{n_R})^{\top}$,
where $\phi^\alpha_c$ ($\varphi^\alpha_c$) is the amplitude of the $c$-th incoming (outgoing)
channel of region $\alpha=L,R$ ($c=1,\cdots,n_\alpha$). 
We also note that $S^\dagger S=SS^\dagger={\bf 1}_{n_L+n_R}$.
The channels are flux-normalised so that
$|\phi^\alpha_c|^2$ is the probability flux carried by that channel~\cite{Fisher1981PRB}.
Accordingly, $r$ ($r'$) is an $n_L\times n_L$ ($n_R\times n_R$) reflection matrix, and
$t$ ($t'$) is the $n_R\times n_L$ ($n_L\times n_R$) transmission matrix from the left to the right
(from the right to the left) side.
Then, the matrix $t$ has rank at most $\min(n_L,n_R)$, which equals $n_R$ in range 1.
By the rank--nullity theorem applied to its $n_L$ columns, $\dim\ker(t)+\operatorname{rank}(t)=n_L$,
so $\dim\ker(t)\ge n_L-n_R$. 
Owing to this, at least $n_L-n_R$ linearly independent incoming states
$\tilde\phi^L$, each a linear combination of the $\phi^L_c$, should satisfy $t\tilde\phi^L=0$.
For such a state, unitarity gives
$|r\tilde\phi^L|^2+|t\tilde\phi^L|^2=|\tilde\phi^L|^2$, hence $|r\tilde\phi^L|^2=|\tilde\phi^L|^2$, i.e., a perfectly reflected state. Therefore, such a reflected state always exists on the side with
more channels, whenever $n_L\neq n_R$.
From Fig.~\ref{fig1}{\bf a}, one finds that in range 1 (range 2)
the BTG (RTG) side has one perfectly reflected state.
We note that this conclusion does not rest on any assumption about the Hamiltonians of the domains or the profile of the boundary.

\subsection*{Adiabatic confinement of massless fermions}

Here, we show that the massless fermion in ABA stacking domain is the perfectly reflected state shown above. 
In one valley and in the sublattice basis $(|A_1\rangle,|B_1\rangle,|A_2\rangle,|B_2\rangle,|A_3\rangle,|B_3\rangle)$ where $|A(B)_\alpha\rangle$ $(\alpha=1,2,3)$ is an orthonormal $p_z$-orbital basis state at the sublattice $A(B)$ of $\alpha$-th layer, the Schr\"odinger equation of the junction geometry discussed above is given by 
\begin{equation}
H [\lambda(x)]\Psi= \left[H_D
+H_\perp^{\rm ABA}+\lambda(x)\Delta H\right]\Psi=E\Psi,
\label{Eq:Total_Hamil_Junct}
\end{equation}
where
$\lambda(x)=\tfrac{1}{2}[1+\tanh({x}/{l})]$, 
$H_D=\hbar v (k_x\sigma_x+k_y\sigma_y)\otimes{\bf 1}_3$,
$H_\perp^{\rm ABA}=\gamma_1\bigl(|B_1\rangle\langle A_2|+|A_2\rangle\langle B_3|\bigr)+\text{h.c.}$,
$H_\perp^{\rm ABC}=\gamma_1\bigl(|B_1\rangle\langle A_2|+|B_2\rangle\langle A_3|\bigr)+\text{h.c.}$, and
$\Delta H=H_\perp^{\rm ABC}-H_\perp^{\rm ABA}
=\gamma_1\bigl(|B_2\rangle\langle A_3|-|A_2\rangle\langle B_3|\bigr)+\text{h.c.}$.
Then, we confirm that, 
for $\lambda(x=+\infty)=1$,  $H\rightarrow H^{\rm ABC}=H_D+H_\perp^{\rm ABC}$ and, 
for $\lambda(x=-\infty)=0$, 
$H\rightarrow H^{\rm ABA}=H_D+H_\perp^{\rm ABA}$.
We note that
$\Delta H$ contains no momentum and has the coupling between the states only in layer 2 and 3. 

In the basis
$|L_\pm\rangle =(|L_1\rangle\pm |L_3\rangle)/\sqrt2$ for a mirror operator $M_z$ satisfying $M_z|L_\pm\rangle=\pm|L_\pm\rangle$ ($L=A,B$),
the spatially varying part in Eq.~\eqref{Eq:Total_Hamil_Junct} 
is rewritten as
\begin{eqnarray}
\lambda(x)\Delta H&=&-g\lambda(x)\bigl(|B_2\rangle\langle A_-|-|A_2\rangle\langle B_-|\bigr)\nonumber \\
& & +g\lambda(x)\bigl(
|B_2\rangle\langle A_+|-|A_2\rangle\langle B_+|\bigr)
+\text{h.c.},
\label{Eq:mirror_coupling}
\end{eqnarray}
where $g=\gamma_1/\sqrt2\simeq 276$ meV.
Let $P_\pm$ be a project operator on its even (odd) or $+1(-1)$ eigenspaces.  
$H_\perp^{\rm ABA}$ commutes with $M_z$, so in the ABA domain the odd block is exactly
$P_oH^{\rm ABA}P_o=\hbar v (k_x\sigma_x+k_y\sigma_y)$~\cite{Min2008PTPS}.
Inside the wall, $\Delta H$ breaks $M_z$, but Eq.~\eqref{Eq:mirror_coupling} carries at most one
odd index, so 
$P_o\,\Delta H\,P_o=0$. Then, $H[\lambda]$ in Eq.~\eqref{Eq:Total_Hamil_Junct} satisfies
$
P_oH[\lambda]P_o=\hbar v(k_x\sigma_x+k_y\sigma_y),
$
for every $\lambda(x)$ in the junction region of $|x|<l$,
implying that the junction adds no mass on the propagating massless state from BTG region.
Instead, what the junction wall does to the odd mode modifies its kinematics as discussed below.

We note that $\Delta H$ in Eq.~\eqref{Eq:mirror_coupling} couples the odd sector only through the layer 2 with 
$g\lambda(x)$.
Since $k_y$ is
conserved,  
we can rewrite Eq.~\eqref{Eq:Total_Hamil_Junct} with fixed $E$ and $k_y$  as a stationary problem for evolution of states $\Psi$ along $x$,
\begin{equation}
\partial_x\Psi=
i {\mathcal K}(x;E,k_y)\Psi,
\label{Eq:stationary}
\end{equation}
where ${\mathcal K}(x;E,k_y)=
\frac{1}{\hbar v}\,{\mathcal B}\bigl[E-\hbar v k_y({\bf 1}_3\otimes\sigma_y)
-H^{\rm ABA}_\perp-\lambda(x)\Delta H
\bigr]$
and 
${\mathcal B}={\bf 1}_3\otimes\sigma_x$.
Owing to pseudo-unitarity ${\mathcal B} {\mathcal K}={\mathcal K}^\dagger{\mathcal B}$, 
the current ${\mathcal J}\equiv\Psi^\dagger{\mathcal B}\Psi$ is conserved, and a local mode of $u_n$ satisfying
${\mathcal K}u_n=k_nu_n$ carries the flux $j_n=u_n^\dagger{\mathcal B} u_n$~\cite{Mello1988AP, Beenakker1997RMP}. 
Here, $j_n$ is zero for complex $k_n$ and nonzero for a real $k_n$,
that is positive (negative) for a right (left) mover. 

By expanding the state $\Psi$ on the local modes  $u_n(x)$ of eigenstates of $\mathcal K$, $\Psi=\sum_nc_n(x)u_n(x)\exp(i\int^x k_n dx)$, 
each propagating mode can be normalised to unit flux at every $x$. 
Inserting this into Eq.~\eqref{Eq:stationary}, the
$k_n$ terms cancel and projecting with the left eigenvector $u_m^\dagger{\mathcal B}/j_m$ gives the following
coupled-mode equations,
\begin{equation}
\partial_xc_m=-\sum_nD_{mn}\,e^{i\int^x(k_n-k_m)}c_n,
\label{eq:cme}
\end{equation}
where $D_{mn}\equiv u_m^\dagger{\mathcal B}\partial_xu_n/j_m$.
The reformulations of Eq.~\eqref{Eq:Total_Hamil_Junct} to Eqs.~\eqref{Eq:stationary} and~\eqref{eq:cme} can be regarded as a real space generalization ~\cite{Johnson2002PRE, Jang2025JCTC} of the  typical time-dependent non-adiabatic coupled-mode equation~\cite{Tully1990JCP}.

From the eigenvalue equation, ${\mathcal K}u_n=k_nu_n$,
the allowed propagating modes in the range 1 in Fig.~\ref{fig1}{\bf a} can be determined through 
$\det[{\mathcal K}(x)-k_n]=0$. 
The characteristic polynomial of $\mathcal K$ can be compactly written as 
\begin{equation}
    R(u)\equiv u\left[(u-c)^2-2\gamma_1^2\tilde{E}^2\right]-\gamma_1^4 E^2\lambda^2=0,
    \label{Eq:R}
\end{equation}
where $u=\hbar^2v^2|{\bf k}|^2-E^2$,
$c=\gamma_1^2\lambda(1-\lambda)$ and $\tilde E^2=E^2-2\lambda(1-\lambda)\hbar^2v^2k_y^2$.
We note that $\lambda$, $E$ and $k_y$ are varying parameters, satisfying $\hbar v|k_y|<E$ and $0\le \lambda \le 1$. 
When $\lambda=0$ for ABA stacking, the roots of Eq.~\eqref{Eq:R} are $u_o=0$ and $u_e^\pm=\pm\sqrt{2}\gamma_1 E$, which corresponds to massless odd and massive even states, respectively, indicating two right-movers in $x<-l$, $k_o=|E|/\hbar v$ from $u_o=0$ and $k_e\simeq 2^{1/4}\sqrt{\gamma_1|E|}/\hbar v$ from $u_e^+=\sqrt{2}\gamma_1 E$.
When $\lambda=1$, it has only one positive root 
$u_\text{ABC}\simeq \gamma_1^{4/3}E^{2/3}$
that corresponds to the right-mover $k_\text{ABC}\simeq \gamma_1^{2/3}|E|^{1/3}/\hbar v$ in ABC stacking. 
For $0<\lambda<1$, it is straightforward to show that Eq.~\eqref{Eq:R} has only one propagating root of $u_b$ larger than $c+\sqrt{2}\gamma_1\tilde{E}$ 
and that $u_b$ is separated from all the other roots by at least $(\sqrt{2}-\sqrt{6}/3)\gamma_1\tilde{E}$. 
Furthermore, one can confirm that $u_b$ adiabatically connects $u_e^+$ to $u_\text{ABC}$ as $\lambda$ varies from 0 to 1. 
In sharp contrast to this, the propagating mode for odd states ceases to exist 
when $\lambda > (E/\gamma_1)(1-{k_y^2}/(4|{\bf k}|^2))$. 
From these, we can assign a unique right moving state $u_e^+$ and its corresponding $k_e$ 
for adiabatically continued states from the left to side side whereas the odd mode does not have such states. 

Along the right-moving even branch $u^+_e$,
we find
that the current $j_e=(u_e^+)^\dagger{\mathcal B} u_e^+$ obeys
$\partial_xj_e=2j_e\mathrm{Re}\,D_{ee}$.
Then, the unit flux condition at every $x$ therefore forces
$\mathrm{Re}D_{ee}=0$ or $D_{ee}$ is a pure imaginary number. 
The resulting diagonal equation integrates to a pure phase so that 
$
c_e(+\infty)=\exp\left(-\int D_{ee}dx\right)=e^{i\gamma_e}$ 
from Eq.~\eqref{eq:cme}, directly indicating that, to the first order, 
the transmission probability of the even massive fermion across ABA and ABC stacking is $|c_e|^2=1$ at every $k_y$. 
Since there is only one mode in ABC stacking in the energy range 1 of Fig.~\ref{fig1}{\bf a}, there is no chance for the odd sector to be connected adiabatically. 
So, the odd mode is neither transmitted nor converted so that it is reflected as itself with unit
probability, at every $k_y$ in the range.

The only way out for the odd mode is a non-adiabatic coupling to other modes within the junction region of $|x|<l$ for the higher-order process, and its size can be estimated from Eq.~\eqref{eq:cme}. 
By differentiating ${\mathcal K}u_n=k_nu_n$ and projecting the differentiated equation with $u_m^\dagger{\mathcal B}/j_m$, 
then one can obtain analytic expression for $D_{mn}$ in case of $m\neq n$ using
$\partial_x{\mathcal K}=-\frac{d \lambda(x)}{dx}{\mathcal B}\Delta H/\hbar v$, as following,
\begin{equation}
D_{mn}=-\frac{\lambda'(x)}{\hbar vj_m}\frac{\langle u_m|\Delta H|u_n\rangle}{k_n-k_m}. 
\label{eq:Dmn}
\end{equation}
The amplitude leaked onto branch $m$ from $n$ can be approximated as  $\partial_x c_m\simeq- D_{mn}e^{i\int^x(k_n-k_m)dx}$.
The spinors of $u_{m}$ and $u_n$ and the
mismatch $\Delta k=k_n-k_m$ would be quite smoothly varied with $x$. 
So, using Eq.~\eqref{eq:Dmn},  
$c_m\propto \int dx\lambda'(x) e^{i x\Delta k }$. 
The leaked amplitude is then
proportional to the Fourier transform of the stacking gradient 
$\lambda'(x)=\frac{1}{2l}{\rm sech}^2(x/l)$
at the boundary region,
\begin{equation}
I(\Delta k)=\int\lambda'(x)e^{i\Delta k\,x}dx
=\frac{\pi l\Delta k/2}{\sinh(\pi l\Delta k/2)}.
\label{eq:I}
\end{equation}
For two limiting cases, 
$I=1$ for $l\Delta k\ll1$ and $I\simeq\pi l\Delta k\,e^{-\pi l\Delta k/2}$ for
$l\Delta k\gg1$. 
Then, we expect a significant leakage of massless fermion to ABC region for the abrupt boundary, while no transmission through the higher order process regardless of $k_y$ for the smooth one.

\subsection*{Quantized energy and wave functions for ideal triangular confinement}

Here, we derive the quantization rule for a massless Dirac fermion in an equilateral triangle of side length $L_M$ satisfying the BM boundary condition~\cite{Berry1987PRSL}. 
Inside the triangle the particle is described by $H_\triangle=-i\hbar v(\sigma_x\partial_x+\sigma_y\partial_y)$ acting on the spinor $\psi=(\psi_1,\psi_2)^{\top}$.
We note that the valley index is suppressed, since the smooth boundary does not mix valleys. 
As shown in 
Extended Data Fig.~\ref{EFig_triangle}{\bf a}, we take the vertices $A=(0,0)$, $B=L_M(1,0)$, $C=L_M(\tfrac12,\tfrac{\sqrt3}{2})$ and label the sides $AB$, $BC$, $CA$ by $s=0,1,2$. Side $s$ has unit tangent $\hat t_s=(\cos\alpha_s,\sin\alpha_s)$ and outward normal $\hat n_s=(\sin\alpha_s,-\cos\alpha_s)$ with $\alpha_s=2\pi s/3$.

A confining boundary condition must be local, linear and carry no current through the side~\cite{Berry1987PRSL,gaddah2018JphysA,Dietz2021APP}. The normal current is $j_n=v\psi^\dagger(\bm\sigma\cdot\hat n_s)\psi=-2v\,\mathrm{Im}\bigl(e^{-i\alpha_s}\psi_1^*\psi_2\bigr)$, with $\bm\sigma=\hat x\sigma_x+\hat y\sigma_y$, so it vanishes when $\psi_2=e^{i\alpha_s}\psi_1$. On side $s$ the BM condition is therefore $\psi=M_s\psi$ with $M_s=\bm\sigma\cdot\hat t_s$.

Since $H_\triangle^2\psi=-\hbar^2v^2\nabla^2\psi=E^2\psi$, each spinor component $\psi_a$ ($a=1,2$) obeys the Helmholtz equation $(\nabla^2+k^2)\psi_a=0$ with $k=E/\hbar v$, so an eigenstate is a superposition of plane waves of the single modulus $k$. 
For $\mathbf k_\theta=k(\cos\theta,\sin\theta)$ the positive-energy spinor is $\psi(\mathbf r;\theta)=(1,e^{i\theta})^{\top}e^{i\mathbf k_\theta\cdot\mathbf r}$, 
and the eigenfunction is $\Psi=\sum_jc(\theta_j)\psi(\mathbf r;\theta_j)$. 
An arbitrary position on side $s$ can be written as $\mathbf r=\mathbf r_s+\tau\hat t_s$ with $\mathbf r_0=A$, $\mathbf r_1=B$, $\mathbf r_2=C$ and $0\le\tau\le L_M$. 
The BM condition on side $s$ then reads $\sum_jc(\theta_j)\bigl(e^{i\theta_j}-e^{i\alpha_s}\bigr)e^{i\mathbf k_j\cdot\mathbf r}=0$ for every $\tau$. 
From this, one can check that waves with different tangential momenta along the sides are linearly independent in $\tau$ and cannot cancel one another. 
So, only a direction $\theta$ and its mirror image $\bar\theta\equiv2\alpha_s-\theta$ share a tangential momentum, and the BM condition can be applied to those pairs.

As shown in 
Extended Data 
Fig.~\ref{EFig_triangle}{\bf b}, the triangle has a mirror image with respect to each of its three sides, and the set of directions must be closed under the three mirrors $\theta\mapsto2\alpha_s-\theta$. 
Two successive mirrors are rotations by $\pm2\pi/3$, so for a given $\mathbf k_\theta$ the triangular geometry generates three equivalent momenta $\mathbf k_{+,s}$ with angles $\theta_{+,s}\equiv\theta+\alpha_s$ and their mirror partners $\mathbf k_{-,s}$ with angles $\theta_{-,s}\equiv-\theta+\alpha_s$, six plane waves in total (on side $s'$, $\theta_{+,s}$ is paired with $\theta_{-,2s'-s}$, indices mod 3). For each mirror pair $(\theta,\bar\theta)$ on side $s$ the boundary condition yields
\begin{equation}
    \frac{c(\bar\theta)}{c(\theta)}=
    e^{i\mathbf r_s\cdot(\mathbf k_\theta-\mathbf k_{\bar\theta})}
    e^{i(\theta-\alpha_s)}.
    \label{Eq:BM_Triangle}
\end{equation}
From this one notes that a wave running along the side ($\theta=\alpha_s$) is self-paired and unconstrained, whereas a wave running the opposite way ($\theta=\alpha_s+\pi$) is not allowed.

Now, let's apply the BM condition in Eq.~\eqref{Eq:BM_Triangle} on each side $s$ directly.
For sides $AB$ and $CA$, 
${\bf r}_s$ is orthogonal to ${\bf k}_\theta -{\bf k}_{\bar{\theta}}$ and $e^{i\mathbf r_s\cdot(\mathbf k_\theta-{\mathbf k}_{\bar\theta})}=1$. 
On side $AB$,
the mirror image of $\theta_{+,s}$ is $\bar{\theta}=-\theta-\alpha_s=\theta_{-,-s}$, 
giving $\tfrac{c(\theta_{-,-s})}{c(\theta_{+,s})}=e^{i(\theta+\alpha_s)}$.
Here, `$-s$' follows mod-3 convention such that $s=-1(-2)$ converts to $2(1)$.
On side $CA$, 
$\bar{\theta}=\tfrac{2\pi}{3}-\theta-\alpha_s=\theta_{-,1-s}$, 
so 
$\tfrac{c(\theta_{-,1-s})}{c(\theta_{+,s})}=e^{i(\theta+\alpha_s-4\pi/3)}$. 
Taking $c(\theta)=1$, then, we can set coefficients, 
$c(\theta_{+,s})=\omega^{s}$ and $c(\theta_{-,s})=e^{i\theta}\omega^{s}$
where $\omega=e^{2\pi i/3}$.

On side $BC$, $\bar\theta=\tfrac{4\pi}{3}-\theta=\theta_{-,2-s}$.
Then, from Eq.~\eqref{Eq:BM_Triangle} and solutions $c(\theta_{\pm,s})$ obtained above, 
one can find 
$e^{i\mathbf r_1\cdot(\mathbf k_{+,s}-\mathbf k_{-,2-s})}=e^{-i(\theta_{+,s}-\alpha_1)}\tfrac{c(\theta_{-,2-s})}{c(\theta_{+,s})}=\omega^{3(1-s)}=1$. 
Now, plugging ${\bf r}_1 = L_M\hat{x}$, we have a quantization condition, 
$e^{ik_sL_M}=1$ where 
$k_s=\hat{x}\cdot (\mathbf k_{+,s}-\mathbf k_{-,2-s}) 
=\sqrt3k\sin(\theta+{2\pi(s-1)}/{3})$. 
With $(k_x,k_y)=k(\cos\theta,\sin\theta)$, we first set  
$k_1L_M=\sqrt{3}k_y L_M=2\pi n$, and then remaining two conditions, $k_0 L_M=-2\pi m$ and $k_2 L_M = 2\pi(m-n)$
with integers $m$ and $n$, satisfying $\sum_s k_s =0$.
Hence, the allowed momentum with the BM  condition on the equilateral triangle is given by
$
{\bf k}_{mn}=\frac{2\pi}{3L_M}\bigl(2m-n,\sqrt3\,n\bigr). 
$
Then, $|\mathbf k_{mn}|=k_{mn}=\frac{4\pi}{3L_M}\sqrt N$ with $N=m^2-mn+n^2$.
With all these ingredients, one can obtain the quantized energy level confined in the equilateral triangle of side $L_M$,
\begin{equation}
   E_{mn}=\hbar vk_{mn}= 
   \frac{4\pi\hbar v}{3L_M}\sqrt{m^2-mn+n^2},
\end{equation}
that corresponds to the bare energy 
$\varepsilon_{mn}$ in Eq.~\eqref{eq:Triangle_Energy} of the main text. 

As shown in 
Extended Data Fig.~\ref{EFig_triangle}{\bf c}, $\mathbf k_{mn}$ runs over the triangular lattice spanned by the reciprocal lattice vectors $\mathbf b_1$ and $\mathbf b_2$ of the periodic lattice generated by the mirror operations on the triangle $ABC$. 
Scaled by $\tfrac{3L_M}{4\pi}$, the lattice points $m\mathbf b_1+n\mathbf b_2$ coincide exactly with the Eisenstein integers $z_{mn}=m+n\omega$ in the complex plane. 
The quantized energy also can be expressed as a norm of $z_{mn}$, 
$N[z_{mn}]=N$, 
where $N[x+iy]=x^2+y^2$ for $x,y\in\mathbb{R}$. 
Among these points, distinct states correspond to one lattice vector per $60^\circ$ wedge, $m>n\ge0$, as shown in Fig.~\ref{fig4}{\bf a}, 
because each of six waves occupies a corresponding 1/6 of entire lattices.

We can obtain the eigenstates as a linear combination of six plane waves for given $m$ and $n$. 
Using computed coefficients above
and $e^{i\theta}=z_{mn}/\sqrt N$, 
the compact expression for each spinor component can be written as 
\begin{eqnarray}
\psi_1^{mn}&=&\Phi_+(\mathbf k_{mn},{\bf r})
+\frac{z_{mn}}{\sqrt N}\Phi_+(\mathcal M\mathbf k_{mn},{\bf r}),\nonumber \\
\psi_2^{mn}&=&\frac{z_{mn}}{\sqrt N}\Phi_-(\mathbf k_{mn},{\bf r})
+\Phi_-(\mathcal M\mathbf k_{mn},{\bf r}),\nonumber
\end{eqnarray}
where 
$\Phi_\pm(\mathbf k,\mathbf r)
=\sum_{s=0}^{2} \omega^{\pm s}
\exp\left[{i(\mathcal C_3^{s}\mathbf k)\cdot\mathbf r}\right]$.
Here, $\mathcal C_3$ is the rotation by $2\pi/3$ and $\mathcal M$ the reflection in the $x$ axis (side $AB$).  
For a given $k_{mn}$,
$\mathcal C_3\mathbf k_{mn}=\mathbf k_{-n,m-n}$, $\mathcal C_3^2\mathbf k_{mn}=
\mathbf k_{n-m,-m}$ and $\mathcal M\mathbf k_{mn}=\mathbf k_{m-n,-n}$. 
Then, with a eigenspinor $\Psi_{mn}=(\psi^{mn}_1,\psi^{mn}_2)^\top$, $\rho_{mn}=|\Psi_{mn}|^2$ and 
the local density for an energy $E_N=\tfrac{4\pi\hbar v}{3L_M}\sqrt{N}$ is
\begin{equation}
    \rho_N(\mathbf r)=
    \sum_{\{m,n\}}\rho_{mn}
    =
    \frac{1}{6S}\sum_{\{m,n\}}\bigl[9-|S_{mn}(\mathbf r)|^2\bigr],
    \label{Eq:rho_N}
\end{equation}
where $S=\sqrt{3}L_M^2/4$, 
$S_{mn}(\mathbf r)=\sum_{s=0}^{2}
\exp\left[{i(\mathcal C_3^{s}\mathbf k)\cdot\mathbf r}\right]$
and $\{m,n\}$ indicates integers $m$ and $n$ satisfying 
$|z_{mn}|=\sqrt{N}$.

\subsection*{Density of states of ideal triangular confinement of massless Dirac fermions}

We first define
$r(N)\equiv\#\{(m,n)\in\mathbb{Z}^2\, |\,|z_{mn}|^2=N\}$,
the number of $z_{mn}=m+ne^{2\pi i/3}$ whose norm is equal to $N$.
Then, the total number ${\mathcal N}(\varepsilon)$ of $z_{mn}$ inside the circle of radius $\varepsilon$ is given by 
${\mathcal N}(\varepsilon)
=\sum_{0 \le N\le \varepsilon^2}r(N)
=\sum_{m,n}\Theta(\varepsilon-|z_{mn}|)$
where $\Theta(x)=1~(0)$ for $x\ge0~(x<0)$.
Then, the total number $n_{\triangle}(\varepsilon)$ 
of $z_{mn}$ in the domain ${\mathcal D}$ shown in Fig.~\ref{fig4}{\bf a} 
can be written as 
$n_{\triangle}(\varepsilon)
=({\mathcal N}(\varepsilon)-1)/6$.

Using Proposition 4.7 in Ref.~\cite{Cox1989}  for the prime factorization of $z_{mn}$, 
the norm can be expressed as a product of primes,
$
N=3^{\alpha_0}\prod_{p\equiv 1(3)}p^{\alpha_p}\prod_{q\equiv2(3)}q^{2\beta_q},
$
where $\alpha_0$, $\alpha_p$ and $\beta_q$ are non-negative integers 
and $p\equiv1(3)$ and $q\equiv2(3)$ denotes primes with residues 1
and 2 modulo 3, respectively~\cite{Pinsky1980SIAM}. 
Applying Proposition 4 in Ref.~\cite{Pinsky1980SIAM} then gives
$r(N)=6\prod_{p\equiv1(3)}(\alpha_p+1)$. 
As an illustration, 
for
$N=1,3,4,7$ and $9$, 
the exponents 
$(\alpha_0,\alpha_p,\beta_q)$ take values $(0,0,0)$, $(1,0,0)$, $(0,0,1)$, $(0,1,0)$ and $(2,0,0)$ yielding $r(1)=r(3)=r(4)=6$, $r(7)=12$ and $r(9)=6$.

To compute ${\mathcal N}(\varepsilon)$, we apply the Poisson summation formula to the dual space $\Lambda^*$ spanned by ${\boldsymbol \xi}=m{\bf p}_1^*+n{\bf p}_2^*$, where ${\bf p}_1^*=(1,-1/\sqrt{3})$ and ${\bf p}_2^*=(0,2/\sqrt{3})$
with $|{\boldsymbol \xi}|^2=\frac{4}{3}N[z_{mn}]=\frac{4}{3}N$. 
Here $\Lambda^*$ is the dual lattice of the triangular lattice $\Lambda$ span by ${\bf p}_1 =(1,0)$ and ${\bf p}_2=(1/2,\sqrt{3}/2)$ satisfying ${\bf p}_i\cdot {\bf p}_j^*=\delta_{ij}$ ($i,j=1,2$), which is equivalent to that 
generated by $z_{mn}$.
Using ${\mathcal N}(\varepsilon)=\frac{1}{\sqrt{3}/2}\sum_{{\boldsymbol{\xi}\in\Lambda^*}}\int^{\varepsilon}_0 \rho d\rho \int_0^{2\pi}d\theta~e^{-2\pi i\rho|{\boldsymbol{\xi}}|\cos\theta}$, 
one obtain 
\begin{equation}
    {\mathcal N}(\varepsilon)=\frac{2}{\sqrt{3}}\pi\varepsilon^2
    +\varepsilon\sum_{N\ge 1}\frac{r(N)}{\sqrt{N}}J_1(2\pi\varepsilon\xi_N),
    \label{Eq:Bessel}
\end{equation}
where $\xi_N=|\boldsymbol{\xi}|=\frac{2}{\sqrt{3}}\sqrt{N}$
and $J_1(x)$ is the first-order Bessel function of the first kind.
Here we note that 
a convolution with the Gaussian broadening kernel
$\frac{1}{{\sqrt{2\pi}\sigma}}e^{-\varepsilon^2/2\sigma^2}$ 
introduces a damping factor $e^{-2\pi^2\sigma^2\xi_N^2}$ 
for each Bessel term.
Owing to its rapid decay with increasing $N$, the contribution from $N=1$ is enough. 
Then, with 
$d_\triangle (\varepsilon)=\partial_\varepsilon
{n}_\triangle(\varepsilon)=
\partial_\varepsilon{\mathcal N}(\varepsilon)/6$, 
one obtains Eq.~\eqref{eq:DOS}.

\subsection*{Leakage of lower energy confined states near $E_F$} 

As discussed in the main text, the ABB stacking on the corner has different interlayer interaction compared with BTG and RTG. 
From Slater-Koster parameters adopted in our work (see Methods), 
the vertical interlayer hopping for direct overlap between atoms, $\gamma_{BB}$, is given as
$\gamma_{BB}/\gamma_1=\exp[-(q_\sigma/a_\sigma)\Delta d]=e^{-2.218\,\Delta d}$, $\Delta d$ in \AA. 
From the the relaxed height corrugation across the
corner shown in Extended Data Figs.~\ref{Efig_ABB}{\bf c}
and ${\bf e}$, 
one can estimate $\gamma_{BB}\simeq 244$ meV. 
With this, we compute the energy bands of pristine ABB stacked trilayer and associated
weight expectation $\langle P_o\rangle$ of odd state projector $P_o$ where 
$\langle P_o\rangle=\langle\psi^{\rm ABB}_{n\mathbf k}|P_o|\psi^{\rm ABB}_{n\mathbf k}\rangle$,
$P_o=|A_-\rangle\langle A_-|+|B_-\rangle\langle B_-|$
and $\psi^{\rm ABB}_{n\mathbf k}$ is a Bloch state of pristine ABB stacking for a given band $n$ at momentum $k$.
From Extended Data Fig.~\ref{Efig_ABB}, one can estimate energy-dependent leakage from the triangular domain to the other as discussed in the main text. 

For a leakage from $n=0$ modes in Fig.~\ref{fig4} {\bf a},
we consider the mirror-odd propagating state in $x<-l$, that is tangent to the wall in $|x|<l$. The boundary geometry is the same as discussed above. Then, the states has $k_x=0$ and $E=\hbar vk_y$, 
and its spinor is given by $|o\rangle=\tfrac{1}{\sqrt2}(|A_-\rangle+i|B_-\rangle)$. 
The coupling $\lambda\Delta H$ in Eq.~\eqref{Eq:mirror_coupling} maps it onto a single normalized, mirror-even state on the layer 2, $|p\rangle=\tfrac{1}{\sqrt2}(-|B_2\rangle+i|A_2\rangle)$, with element $\langle p|\lambda\Delta H|o\rangle=g\lambda$. 
So, they can be regarded as degenerate states with an effective Hamiltonian, 
$H_\text{eff}=E+g\lambda\sigma_x$ for $(|o\rangle, |p\rangle)^T$. 
Under $H_\text{eff}$, the odd state will evolves as $(\cos\vartheta,-i\sin\vartheta)^\top$. 
So, assuming that massless particle moves at speed $v$ with arbitrary momentum along a path $s$, then
$\vartheta=g\lambda(x)s/\hbar v$. 
Near a side the $n=0$ odd state is spread over a transverse width $L_\rho\sim L_M$, each slice $dx/L_\rho$ runs a path $L_\rho$. 
The rotation of the state can be the density-weighted average, 
$\delta=\int_{-\infty}^{0}\frac{d x}{L_\rho}\,\frac{g\lambda(x)}{\hbar v}\,L_\rho
= \frac{\gamma_1 l \ln2}{2\sqrt2\,\hbar v} \simeq 0.330\pi. 
$
Then, this evolution estimation gives $A_w=\cos^2\delta=0.259$. 

\subsection*{Self-energy corrections to confined massless states}

Let's split the Hilbert space into $P$ for the odd modes $\psi_P=\Psi_{mn}$ for the closed triangle with $H_\triangle\Psi_{mn}=\varepsilon_{nm}\Psi_{mn}$ 
and $Q$ for leaked states ($H_Q\psi_Q=\varepsilon_Q\psi_Q$) that $\Psi_{mn}$ couples to
with a coupling $V$. 
Here, $\varepsilon_{nm}=\varepsilon_0|z_{mn}|$ and 
$\varepsilon_0=\tfrac{4\pi\hbar v}{3L_M}$.
Writing
$
    H=\bigl(\begin{smallmatrix}H_\triangle & {V}\\ 
{V}^\dagger & H_Q\end{smallmatrix}\bigr)\nonumber 
$ 
for $(\psi_P, \psi_Q)^\top$,
and eliminating $Q$ based on a standard  procedure~\cite{Min2008PTPS}, 
\begin{equation}
\left[H_\triangle+\Sigma(E)\right]\psi_P
=E\psi_P, 
\nonumber
\end{equation}
where a self energy is given by $\Sigma(E)=V(E-H_Q+i0)^{-1}V^\dagger$.
Here for $\Sigma(E)\equiv\Delta(E)+i\Gamma(E)$, we assume the affine form without mode-coupling such that
$\Delta(E)\simeq \Delta({E}_0)+(E- E_0)\partial_E\Delta(E)|_{E=E_0}$
for an energy $E_0 >E_c$. 
With $\alpha=-\partial_E\Delta(E)|_{E=E_0}$
and $\varepsilon_c=\Delta({E}_0)- E_0\,\partial_E\Delta(E)|_{E=E_0}$, 
we can obtain 
$E_{mn}\simeq Z\varepsilon_0|z_{mn}|+E_c$
where $Z=\tfrac{1}{1+\alpha}$
and $E_c=Z\varepsilon_c$
as shown in Eq.~\eqref{eq:Triangle_Energy} and 
to match the simulated data, we set $Z\simeq 0.84$.

To account the boundary leakage for $n=0$ modes in DOS, we  define 
${\mathcal B}(\varepsilon)=\#\{(m,0)\mid 0 < m\le\varepsilon\}=\sum_{m\ge 1}\Theta(\varepsilon-m)$. 
For ${\mathcal B}(\varepsilon)$, 
an analogues calculation to ${\mathcal N}(\varepsilon)$ in Eq.~\eqref{Eq:Bessel} gives
${\mathcal B}(\varepsilon)=\varepsilon+\frac{1}{\pi}\sum_{n\ge 1}\frac{\sin(2\pi n\varepsilon)}{n}$.
We note that 
a Gaussian convolution
$\frac{1}{{\sqrt{2\pi}\sigma}}e^{-\varepsilon^2/2\sigma^2}$ 
introduces a damping $e^{-2\pi^2\sigma^2n^2}$  for 
$\sin(2\pi n\varepsilon)$ 
when $\varepsilon\gg\sigma$.
Owing to their rapid decays with increasing $n$, the contributions from $n=1$ are enough. 
So, the DOS contribution from the modes along the walls is 
$b_\text{wall}(\varepsilon)=\partial_\varepsilon{\mathcal B}(\varepsilon)\simeq 1+2e^{-2\pi^2\sigma^2}\cos(2\pi\varepsilon)$.
Then, considering a renormalization factor $Z$, the total DOS $D_\triangle$ can be written as 
$D_\triangle = Z(d_\triangle-d_\text{wall})+ZA_w d_\text{wall}$ as can be seen in Eq.~\eqref{Eq:ReDOS}. 

By extracting the oscillating parts in $D_\triangle$ and compare them with pristine DOS of ABA stacking, one can obtain the normalized deviation shown in Fig.~\ref{fig4}{\bf c} as follows, 
\begin{equation}
  \Delta\rho(E)=100\times\frac{D_{\rm osc}(E)}{D_e(E)+D_o(E)}~\text{in}~\%,  
\end{equation}
where
$D_{\rm osc}(E)=\frac{1}{S\varepsilon_0}
\left[
d_{\rm bulk}(\varepsilon)
+(A_w-1)\,(b_{\rm wall}(\varepsilon)-1)\right]$ is the oscillatory part in $D_\triangle$,
$\varepsilon=\tfrac{E-E_c}{Z\varepsilon_0}$,
$S=\frac{\sqrt{3}L_M^2}{4}$, and
$d_\text{bulk}(\varepsilon)=(2\pi\varepsilon)/(3\sqrt3)e^{-8\pi^2\sigma^2/{3}}
    J_0({4\pi\varepsilon}/{\sqrt{3}})$.
Here, 
$D_e(E)=\tfrac{\sqrt{2}\gamma_1}{4\pi\hbar^2 v^2}$ and $D_o(E)=\frac{E}{2\pi\hbar^2 v^2}$ are 
the pristine DOSs of even and odd modes in BTG, respectively.

\section*{Acknowledgments}
We thank fruitful discussions with Mikito Koshino. 
K.Y. was supported by Korea Institute for Advanced Study (KIAS) individual grant (CG092502). 
C.P. was supported by the National Research Foundation of Korea (NRF) grant funded by the Korea government (MSIT and MOE) (No. RS-2025-16063688), and the Basic Science Research Program (No. RS-2024-00349573) through NRF, funded by the MSIT of the Republic of Korea.
Y-W.S. was supported by the KIAS individual Grant No. (CG031509)
and the NRF grant funded by the Korea government (MSIT) (No. RS-2026-25508298).  Computations were supported by CAC of KIAS.

\bibliography{confinement}

@article{andrei2021NRM,
  author    = {Andrei, Eva Y. and Efetov, Dmitri K. and Jarillo-Herrero, Pablo and MacDonald, Allan H. and Mak, Kin Fai and Senthil, T. and Tutuc, Emanuel and Yazdani, Ali and Young, Andrea F.},
  title     = {{The marvels of moir{\'{e}} materials}},
  journal   = {Nat. Rev. Mater.},
  volume    = {6},
  number    = {3},
  pages     = {201--206},
  year      = {2021},
  month     = {Mar},
  doi       = {10.1038/s41578-021-00284-1}
}

@article{Ren2025JPCM,
doi = {10.1088/1361-648X/adf6f9},
year = {2025},
month = {aug},
publisher = {IOP Publishing},
volume = {37},
number = {35},
pages = {353001},
author = {Ren, Wei and Zhu, Ziyan and Zhang, Xi and Luskin, Mitchell and Wang, Ke},
title = {Review: moir\'e-of-moir\'e superlattice in twisted trilayer graphene},
journal = {J. Phys.: Condensed Matter}
}

@article{Kennes2021NatPhys,
  author    = {Kennes, Dante M. and Claassen, Martin and Xian, Lede and Georges, Antoine and Millis, Andrew J. and Hone, James and Dean, Cory R. and Basov, D. N. and Pasupathy, Abhay and Rubio, Angel},
  title     = {Moir{\'e} heterostructures as a condensed-matter quantum simulator},
  journal   = {Nat. Phys.},
  volume    = {17},
  number    = {2},
  pages     = {155--163},
  year      = {2021},
  month     = {Feb},
  publisher = {Nature Publishing Group},
  doi       = {10.1038/s41567-020-01154-3},
  url       = {https://doi.org/10.1038/s41567-020-01154-3}
}

@article{Levendorf2012Nature,
  author        = {Levendorf, Mark P. and Kim, Cheol-Joo and Brown, Lola and Huang, Pinshane Y. and Havener, Robin W. and Muller, David A. and Park, Jiwoong},
  title         = {{Graphene and boron nitride lateral heterostructures for atomically thin circuitry}},
  journal       = {Nature},
  year          = {2012},
  volume        = {488},
  number        = {7413},
  pages         = {627--632},
  doi           = {10.1038/nature11408}
}

@article{Huang2014NM,
  author        = {Huang, Chunming and Wu, Sanfeng and Sanchez, Ana M. and Peters, Jonathan J. P. and Beanland, Richard and Ross, Jason S. and Rivera, Pasqual and Yao, Wang and Cobden, David H. and Xu, Xiaodong},
  title         = {{Lateral heterojunctions within monolayer MoSe$_2$--WSe$_2$ semiconductors}},
  journal       = {Nature Materials},
  year          = {2014},
  volume        = {13},
  number        = {12},
  pages         = {1096--1101},
  doi           = {10.1038/nmat4064}
}

@article{Li2015Science,
  author        = {Li, Ming-Yang and Shi, Yumeng and Cheng, Chia-Chin and Lu, Li-Syuan and Lin, Yung-Chang and Tang, Hao-Lin and Tsai, Meng-Lin and Chu, Chih-Wei and Wei, Kung-Hwa and He, Jr-Hau and Chang, Wen-Hao and Suenaga, Kazu and Li, Lain-Jong},
  title         = {{Epitaxial growth of a monolayer WSe$_2$-MoS$_2$ lateral p-n junction with an atomically sharp interface}},
  journal       = {Science},
  year          = {2015},
  volume        = {349},
  number        = {6247},
  pages         = {524--528},
  doi           = {10.1126/science.aab4097}
}

@article{Leong2018JACS,
  author        = {Leong, Wei Sun and Ji, Qingqing and Mao, Nannan and Han, Yimo and Wang, Haozhe and Goodman, Aaron J. and Vignon, Antoine and Su, Cong and Guo, Yunfan and Shen, Pin-Chun and Gao, Zhenfei and Muller, David A. and Tisdale, William A. and Kong, Jing},
  title         = {{Synthetic Lateral Metal-Semiconductor Heterostructures of Transition Metal Disulfides}},
  journal       = {J. Am. Chem. Soc.},
  year          = {2018},
  volume        = {140},
  number        = {39},
  pages         = {12354--12358},
  doi           = {10.1021/jacs.8b07806}
}

@article{Nakanishi2010PRB,
  author        = {Nakanishi, Takeshi and Koshino, Mikito and Ando, Tsuneya},
  title         = {{Transmission through a boundary between monolayer and bilayer graphene}},
  journal       = {Physical Review B},
  year          = {2010},
  volume        = {82},
  number        = {12},
  pages         = {125428},
  doi           = {10.1103/PhysRevB.82.125428}
}

@article{Ando1998JPSJ,
author = {Ando ,Tsuneya and Nakanishi ,Takeshi and Saito ,Riichiro},
title = {Berry's Phase and Absence of Back Scattering    in Carbon Nanotubes},
journal = {J. Phys. Soc. Jpn.},
volume = {67},
number = {8},
pages = {2857-2862},
year = {1998},
doi = {10.1143/JPSJ.67.2857}
}

@article{Katsnelson2006NatPhys,
  title={Chiral tunnelling and the {Klein} paradox in graphene},
  author={Katsnelson, M. I. and Novoselov, K. S. and Geim, A. K.},
  journal={Nat. Phys.},
  volume={2},
  number={9},
  pages={620--625},
  year={2006},
  publisher={Nature Publishing Group},
  doi={10.1038/nphys384}
}

@article{Ponomarenko2008Science,
  author  = {Ponomarenko, L. A. and Schedin, F. and Katsnelson, M. I. and Yang, R. and Hill, E. W. and Novoselov, K. S. and Geim, A. K.},
  title   = {Chaotic {Dirac} billiard in graphene quantum dots},
  journal = {Science},
  year    = {2008},
  volume  = {320},
  number  = {5874},
  pages   = {356--358},
  doi     = {10.1126/science.1154663}
}

@article{Berry1987PRSL,
  author  = {Berry, M. V. and Mondragon, R. J.},
  title   = {Neutrino billiards: time-reversal symmetry-breaking without magnetic fields},
  journal = {Proc. R. Soc. Lond. A},
  volume  = {412},
  number  = {1842},
  pages   = {53--74},
  year    = {1987},
  doi     = {10.1098/rspa.1987.0080}
}

@article{Chodos1974PRD,
  author  = {Chodos, A. and Jaffe, R. L. and Johnson, K. and Thorn, C. B. and Weisskopf, V. F.},
  title   = {New extended model of hadrons},
  journal = {Phys. Rev. D},
  volume  = {9},
  pages   = {3471--3495},
  year    = {1974},
  doi     = {10.1103/PhysRevD.9.3471}
}

@article{DeMartino2007PRL,
  author  = {De Martino, A. and Dell'Anna, L. and Egger, R.},
  title   = {Magnetic confinement of massless {Dirac} fermions in graphene},
  journal = {Phys. Rev. Lett.},
  year    = {2007},
  volume  = {98},
  number  = {6},
  pages   = {066802},
  doi     = {10.1103/PhysRevLett.98.066802}
}

@article{zhu2020prl,
	title = {Twisted Trilayer Graphene: A Precisely Tunable Platform for Correlated Electrons},
	volume = {125},
	issn = {0031-9007, 1079-7114},
	url = {https://link.aps.org/doi/10.1103/PhysRevLett.125.116404},
	doi = {10.1103/PhysRevLett.125.116404},
	shorttitle = {Twisted Trilayer Graphene},
	pages = {116404},
	number = {11},
	journaltitle = {Physical Review Letters},
	shortjournal = {Phys. Rev. Lett.},
	journal = {Phys. Rev. Lett.},
	author = {Zhu, Ziyan and Carr, Stephen and Massatt, Daniel and Luskin, Mitchell and Kaxiras, Efthimios},
	urldate = {2026-03-04},
	date = {2020-09-11},
	year = {2020},
	langid = {english},
}

@article{Nakatsuji2023PRX,
  title = {Multiscale Lattice Relaxation in General Twisted Trilayer Graphenes},
  author = {Nakatsuji, Naoto and Kawakami, Takuto and Koshino, Mikito},
  journal = {Phys. Rev. X},
  volume = {13},
  issue = {4},
  pages = {041007},
  numpages = {20},
  year = {2023},
  month = {Oct},
  doi = {10.1103/PhysRevX.13.041007},
  publisher = {American Physical Society}
}

@article{Park2025Nature,
	title = {Unconventional domain tessellations in moir\'{e}-of-moir\'{e} lattices},
	volume = {641},
	issn = {0028-0836, 1476-4687},
	url = {https://www.nature.com/articles/s41586-025-08932-0},
	doi = {10.1038/s41586-025-08932-0},
	pages = {896--903},
	number = {8064},
	journaltitle = {Nature},
	journal = {Nature},
	shortjournal = {Nature},
	author = {Park, Daesung and Park, Changwon and Yananose, Kunihiro and Ko, Eunjung and Kim, Byunghyun and Engelke, Rebecca and Zhang, Xi and Davydov, Konstantin and Green, Matthew and Kim, Hyun-Mi and Park, Sang Hwa and Lee, Jae Heon and Kim, Seul-Gi and Kim, Hyeongkeun and Watanabe, Kenji and Taniguchi, Takashi and Yang, Sang Mo and Wang, Ke and Kim, Philip and Son, Young-Woo and Yoo, Hyobin},
	urldate = {2025-05-27},
	date = {2025-05-22},
	  year = {2025},
	langid = {english},
}

@article{latil2006prl,
	title = {Charge Carriers in Few-Layer Graphene Films},
	volume = {97},
	doi = {10.1103/PhysRevLett.97.036803},
	pages = {036803},
	number = {3},
    journal = {Phys. Rev. Lett.},
	author = {Latil, Sylvain and Henrard, Luc},
    year = {2006}
}

@article{Min2008PTPS,
    author = {Min, Hongki and MacDonald, Allan H.},
    title = {Electronic Structure of Multilayer Graphene},
    journal = {Prog. Theor. Phys. Supple.},
    volume = {176},
    pages = {227-252},
    year = {2008},
    month = {06},
    doi = {10.1143/PTPS.176.227}
}

@article{Park2023NatComm,
	title = {Condensation of preformed charge density waves in kagome metals},
	volume = {14},
	number = {1},
	urldate = {2025-02-06},
	journal = {Nat. Commun.},
	author = {Park, Changwon and Son, Young-Woo},
	month = nov,
	year = {2023},
	pages = {7309},
    doi ={10.1038/s41467-023-43170-w}
}

@article{yananose2025hBN,
  title = {Super moir\'e domain tessellations, sliding ferroelectricity, and reconfigurable quantum dot arrays in twisted trilayer hexagonal boron nitride},
  author = {Yananose, Kunihiro and Park, Changwon and Son, Young-Woo},
  journal = {Phys. Rev. B},
  volume = {113},
  issue = {20},
  pages = {205407},
  numpages = {23},
  year = {2026},
  month = {May},
  publisher = {American Physical Society},
  doi = {10.1103/yfqy-6235}
}

@article{Pinsky1980SIAM,
  author  = {Pinsky, M. A.},
  title   = {The eigenvalues of an equilateral triangle},
  journal = {SIAM J. Math. Anal.},
  volume  = {11},
  number  = {5},
  pages   = {819--827},
  year    = {1980},
  doi     = {10.1137/0511073}
}

@article{Itzykson1986JPA,
  author  = {Itzykson, C. and Luck, J.-M.},
  title   = {Arithmetical degeneracies in simple quantum systems},
  journal = {J. Phys. A: Math. Gen.},
  volume  = {19},
  number  = {2},
  pages   = {211--239},
  year    = {1986},
  doi     = {10.1088/0305-4470/19/2/017}
}

@book{Cox1989,
  author    = {Cox, David A.},
  title     = {Primes of the Form $x^2 + ny^2$: {Fermat}, Class Field Theory, and Complex Multiplication},
  year      = {1989},
  publisher = {John Wiley \& Sons},
  address   = {New York},
  series    = {Pure and Applied Mathematics: A Wiley Series of Texts, Monographs and Tracts},
    pages = {77},
  isbn      = {978-0-471-50654-6}
}

@article{Glazman1988JETP,
  author    = {Glazman, L. I. and Lesovik, G. B. and Khmel'nitskii, D. E. and Shekhter, R. I.},
  title     = {Reflectionless quantum transport and fundamental ballistic-resistance steps in microscopic constrictions},
  journal   = {JETP Lett.},
  volume    = {48},
  number    = {4},
  pages     = {238--241},
  year      = {1988},
  url = {http://jetpletters.ru/ps/0/article_16696.shtml}
}

@article{Banin1999Nature,
  title={Identification of atomic-like electronic states in indium arsenide nanocrystal quantum dots},
  author={Banin, Uri and Cao, YunWei and Katz, David and Millo, Oded},
  journal={Nature},
  volume={400},
  number={6744},
  pages={542--544},
  year={1999},
  publisher={Nature Publishing Group},
  doi={10.1038/22979}
}

@book{Lame1852,
  author    = {Lam{\'e}, Gabriel},
  title     = {Le{\c{c}}ons sur la th{\'e}orie math{\'e}matique de l'{\'e}lasticit{\'e} des corps solides},
  publisher = {Bachelier},
  address   = {Paris},
  pages   ={131--136},
  year      = {1852}
}

@book{Mathews1970,
  title={Mathematical Methods of Physics},
  author={Mathews, Jon and Walker, Robert L.},
  edition={2nd},
  year={1970},
  publisher={W. A. Benjamin},
  pages = {237--239},
  address={New York},
  isbn={978-0805370027}
}

@article{Krishnamurthy1982JPA,
  author  = {Krishnamurthy, H. R. and Mani, H. S. and Verma, H. C.},
  title   = {Exact solution of the {Schr\"{o}dinger} equation for a particle in a tetrahedral box},
  journal = {J. Phys. A},
  volume  = {15},
  number  = {7},
  pages   = {2131--2137},
  year    = {1982},
  doi     = {10.1088/0305-4470/15/7/024}
}

@article{mccann2004jphyscond,
	title = {Symmetry of boundary conditions of the Dirac equation for electrons in carbon nanotubes},
	volume = {16},
	doi = {10.1088/0953-8984/16/13/016},
	pages = {2371--2379},
	number = {13},
	journal = {J. Phys.: Condens. Matter},
	author = {{McCann}, Edward and Fal'ko, Vladimir I},
	year = {2004},
}

@article{Brey2006PRB,
  author  = {Brey, L. and Fertig, H. A.},
  title   = {Electronic states of graphene nanoribbons studied with the {Dirac} equation},
  journal = {Phys. Rev. B},
  volume  = {73},
  pages   = {235411},
  year    = {2006},
  doi     = {10.1103/PhysRevB.73.235411}
}

@article{Dietz2021APP,
  author  = {Dietz, B.},
  title   = {Relativistic Quantum Billiards with Threefold Rotational Symmetry: Exact, Symmetry-Projected Solutions for the Equilateral Neutrino Billiard},
  journal = {Acta Phys. Pol. A},
  volume  = {140},
  number  = {6},
  pages   = {473--485},
  year    = {2021},
  doi     = {10.12693/APhysPolA.140.473}
}

@article{gaddah2018JphysA,
	title = {Exact solutions to the Dirac equation for equilateral triangular billiard systems},
	volume = {51},
	doi = {10.1088/1751-8121/aad742},
	pages = {385304},
	number = {38},
	journal = {J. Phys. A: Math. Theor.},
	author = {Gaddah, Wajdi A},
	year = {2018}
}

@article{Koshino2013PRB2,
  title = {Multilayer graphenes with mixed stacking structure: Interplay of Bernal and rhombohedral stacking},
  author = {Koshino, Mikito and McCann, Edward},
  journal = {Phys. Rev. B},
  volume = {87},
  issue = {4},
  pages = {045420},
  numpages = {12},
  year = {2013},
  month = {Jan},
  publisher = {American Physical Society},
  doi = {10.1103/PhysRevB.87.045420}
}

@article{Yankowitz2014NatMater,
  author  = {Yankowitz, Matthew and Wang, Joel I-Jan and Birdwell, A. Glen and Chen, Yu-An and Watanabe, Kenji and Taniguchi, Takashi and Jacquod, Philippe and San-Jose, Pablo and Jarillo-Herrero, Pablo and LeRoy, Brian J.},
  title   = {Electric field control of soliton motion and stacking in trilayer graphene},
  journal = {Nat. Mater.}, 
  volume = {13}, 
  pages = {786--789}, 
  year = {2014},
  doi     = {10.1038/nmat3965}
}

@article{Jiang2018NatNano,
  author  = {Jiang, Lili and Wang, Sheng and Shi, Zhiwen and Jin, Chenhao and Utama, M. Iqbal Bakti and Zhao, Sihan and Shen, Yuen-Ron and Gao, Hong-Jun and Zhang, Guangyu and Wang, Feng},
  title   = {Manipulation of domain-wall solitons in bi- and trilayer graphene},
  journal = {Nat. Nanotechnol.}, 
  volume = {13}, 
  number = {3}, 
  pages = {204--208}, 
  year = {2018},
  doi     = {10.1038/s41565-017-0042-6}
}

@article{Huang2018PR,
title = {Relativistic quantum chaos},
journal = {Phys. Rep.},
volume = {753},
pages = {1-128},
year = {2018},
note = {Relativistic Quantum Chaos},
issn = {0370-1573},
doi = {https://doi.org/10.1016/j.physrep.2018.06.006},
author = {Liang Huang and Hong-Ya Xu and Celso Grebogi and Ying-Cheng Lai}
}

@article{Kolmogorov2005PRB,
  title = {Registry-dependent interlayer potential for graphitic systems},
  author = {Kolmogorov, Aleksey N. and Crespi, Vincent H.},
  journal = {Phys. Rev. B},
  volume = {71},
  issue = {23},
  pages = {235415},
  numpages = {6},
  year = {2005},
  month = {Jun},
  publisher = {American Physical Society},
  doi = {10.1103/PhysRevB.71.235415}
}

@book{lehoucq_arpack_1998,
	address = {Philadelphia},
	title = {{ARPACK} users' guide: solution of large-scale eigenvalue problems with implicitly restarted {Arnoldi} methods},
	shorttitle = {{ARPACK} users' guide},
	publisher = {Society for Industrial and Applied Mathematics},
	author = {Lehoucq, Richard B. and Sorensen, Danny C. and Yang, C. },
	year = {1998},
}

@misc{opencollab_arpack-ng,
	title = {Arpack-ng},
	author = {Opencollab},
    note ={[Accessed: 24-July-2025]},
     howpublished = {\url{https://github.com/opencollab/arpack-ng}}
}

@article{alden2013pnas,
	title = {Strain solitons and topological defects in bilayer graphene},
	volume = {110},
	issn = {0027-8424, 1091-6490},
	url = {https://pnas.org/doi/full/10.1073/pnas.1309394110},
	doi = {10.1073/pnas.1309394110},
	pages = {11256--11260},
	number = {28},
	journaltitle = {Proceedings of the National Academy of Sciences},
	shortjournal = {Proc. Natl. Acad. Sci. U.S.A.},
	journal = {Proc. Natl. Acad. Sci. U.S.A.},
	author = {Alden, Jonathan S. and Tsen, Adam W. and Huang, Pinshane Y. and Hovden, Robert and Brown, Lola and Park, Jiwoong and Muller, David A. and {McEuen}, Paul L.},
	urldate = {2026-04-20},
	date = {2013-07-09},
	year = {2013},
	langid = {english},
}

@article{Fisher1981PRB,
  title = {Relation between conductivity and transmission matrix},
  author = {Fisher, Daniel S. and Lee, Patrick A.},
  journal = {Phys. Rev. B},
  volume = {23},
  issue = {12},
  pages = {6851(R)--6854(R)},
  numpages = {0},
  year = {1981},
  month = {Jun},
  publisher = {American Physical Society},
  doi = {10.1103/PhysRevB.23.6851}
}

@article{Mello1988AP,
title = {Macroscopic approach to multichannel disordered conductors},
journal = {Ann. Phys.},
volume = {181},
number = {2},
pages = {290-317},
year = {1988},
issn = {0003-4916},
doi = {https://doi.org/10.1016/0003-4916(88)90169-8},
author = {P.A Mello and P Pereyra and N Kumar}
}

@article{Beenakker1997RMP,
  title = {Random-matrix theory of quantum transport},
  author = {Beenakker, C. W. J.},
  journal = {Rev. Mod. Phys.},
  volume = {69},
  issue = {3},
  pages = {731--808},
  numpages = {0},
  year = {1997},
  month = {Jul},
  publisher = {American Physical Society},
  doi = {10.1103/RevModPhys.69.731},
  url = {https://link.aps.org/doi/10.1103/RevModPhys.69.731}
}

@article{Johnson2002PRE,
  title = {Adiabatic theorem and continuous coupled-mode theory for efficient taper transitions in photonic crystals},
  author = {Johnson, Steven G. and Bienstman, Peter and Skorobogatiy, M. A. and Ibanescu, Mihai and Lidorikis, Elefterios and Joannopoulos, J. D.},
  journal = {Phys. Rev. E},
  volume = {66},
  issue = {6},
  pages = {066608},
  numpages = {15},
  year = {2002},
  month = {Dec},
  publisher = {American Physical Society},
  doi = {10.1103/PhysRevE.66.066608}
}

@article{Jang2025JCTC,
    author = {Jang, Seogjoo J. and Min, Byeong Ki and Rhee, Young Min},
    title = {Fermi’s
Golden Rule Rate Expression for Transitions
Due to Nonadiabatic Derivative Couplings in the Adiabatic Basis},
    journal = {J. Chem. Theory Comput.},
    volume = {21},
    number = {4},
    pages = {1850-1864},
    year = {2025},
    month = {02},
    issn = {1549-9618},
    doi = {10.1021/acs.jctc.4c00590}
}

@article{Tully1990JCP,
    author = {Tully, John C.},
    title = {Molecular dynamics with electronic transitions},
    journal = {J. Chem. Phys.},
    volume = {93},
    number = {2},
    pages = {1061-1071},
    year = {1990},
    month = {07},
    doi = {10.1063/1.459170}
}

@article{joucken2015carbon,
        title = {Fourier transform analysis of {STM} images of multilayer graphene moiré patterns},
        volume = {83},
        issn = {00086223},
        url = {https://linkinghub.elsevier.com/retrieve/pii/S0008622314011026},
        doi = {10.1016/j.carbon.2014.11.030},
        pages = {48--52},
        journaltitle = {Carbon},
        shortjournal = {Carbon},
        journal = {Carbon},
        author = {Joucken, Fr\'ed\'eric and Frising, Fernande and Sporken, Robert},
        urldate = {2026-06-04},
        date = {2015-03},
        year = {2015},
        langid = {english},
}

@article{yin2016natcomm,
	title = {Direct imaging of topological edge states at a bilayer graphene domain wall},
	volume = {7},
	issn = {2041-1723},
	url = {https://www.nature.com/articles/ncomms11760},
	doi = {10.1038/ncomms11760},
	pages = {11760},
	number = {1},
	journaltitle = {Nature Communications},
	shortjournal = {Nat Commun},
    journal = {Nat Commun.},
	author = {Yin, Long-Jing and Jiang, Hua and Qiao, Jia-Bin and He, Lin},
	urldate = {2026-04-22},
	date = {2016-06-17},
	year = {2016},
	langid = {english},
}

\clearpage
\onecolumngrid

\setcounter{figure}{0}
\setcounter{table}{0}
\renewcommand{\figurename}{Extended Data Fig}
\renewcommand{\tablename}{Extended Data Table}

\begin{figure*}[h!]
\centering
\includegraphics[width=1.00\textwidth]{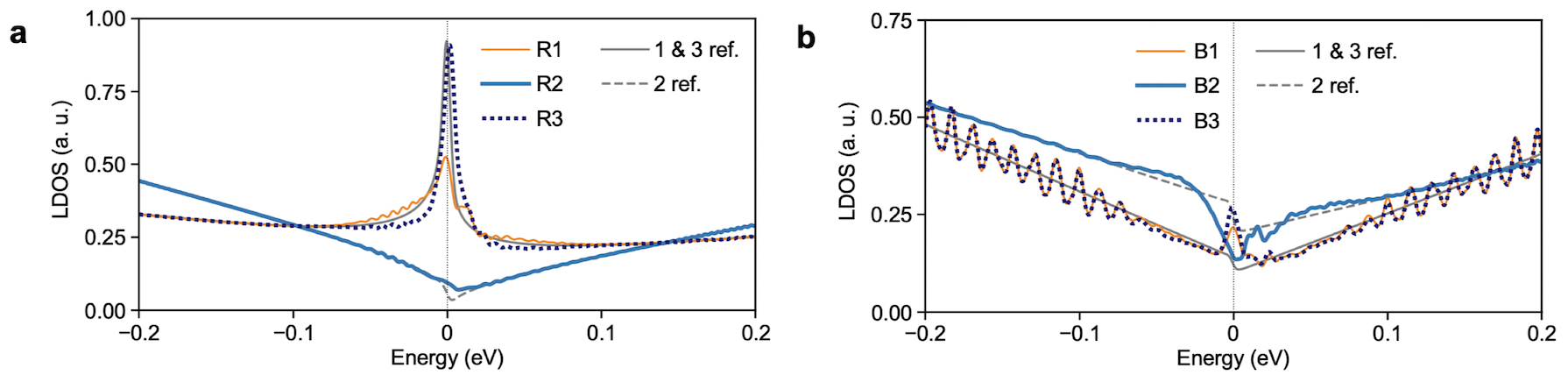}
\caption{
\textbf{Layer-resolved LDOS of the TMB system projected on each domain.}
\textbf{a} The layer-resolved LDOS projected on the rhombohedral domain of the TMB system (black triangle in Fig.~\ref{fig1}\textbf{c}) are shown, where `R1', `R2', and `R3' refer to the LDOS projected on the layer 1, 2, and 3 of the rhombohedral domain, respectively.
Here, layer 1 is the twisted monolayer section of the TMB. 
Reference layer-resolved DOS of the pristine RTG are shown together. Note that in the pristine system, the layer-resolved DOSs of the layer 1 and layer 3 are the same due to a symmetry (labeled as `1 \& 3 ref.'), whereas that of the layer 2 is distinct (`2 ref.')
\textbf{b} The layer-resolved LDOS projected on the Bernal domain of the TMB system (white triangle in Fig.~\ref{fig1}\textbf{c}) are shown where `B1', `B2', and `B3' refer to the LDOS projected on the layer 1, 2, and 3 of the Bernal domain, respectively.
}
\label{EFig_layerLDOS}
\end{figure*}

\begin{figure*}[h!]
\centering
\includegraphics[width=1.00\textwidth]{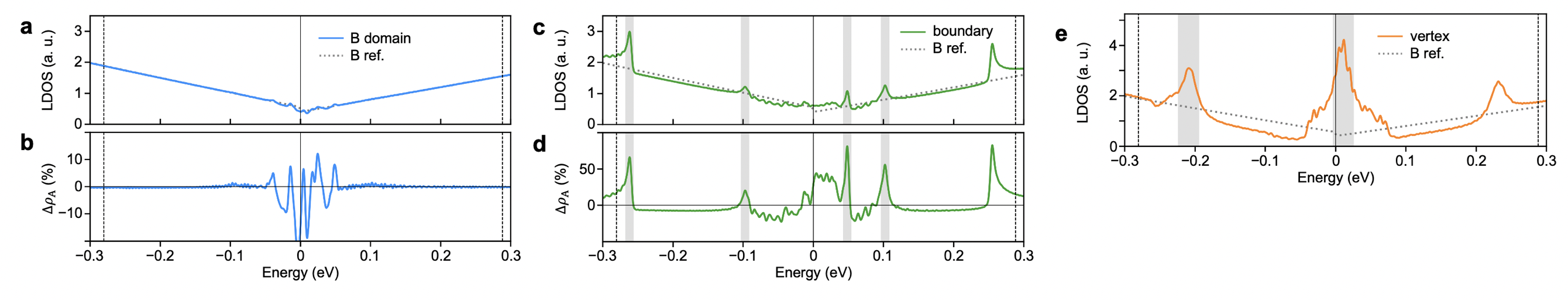}
\caption{
\textbf{LDOS of the ATT system projected on each regions.}
\textbf{a} The LDOS ($\rho_A(E)$) projected on the Bernal (B) domains of the ATT system (white triangles in Fig.~\ref{fig1}\textbf{d}) is shown by a solid line, and the B reference DOS is shown by a gray dotted line. Similarly to the Fig.~\ref{fig2}, the corresponding normalized deviation is shown in \textbf{b}.
\textbf{c} and \textbf{d} The LDOS of the boundary regions of the ATT system (rectangles in Fig.~\ref{fig1}\textbf{d}) and its normalized deviation $\Delta\rho_{A}$. The gray shades are $\pm\sigma$ range around the energy levels corresponding to Extended Data Fig.~\ref{EFig_rho_ATT}\textbf{c-f}. 
\textbf{e} The LDOS of the vertex and the B reference DOS. The gray shades are $\pm\sigma$ range for Extended Data Fig.~\ref{EFig_rho_ATT}\textbf{a,b}. 
}
\label{EFig_LDOS_ATT}
\end{figure*}

\begin{figure*}[h!]
\centering
\includegraphics[width=1.00\textwidth]{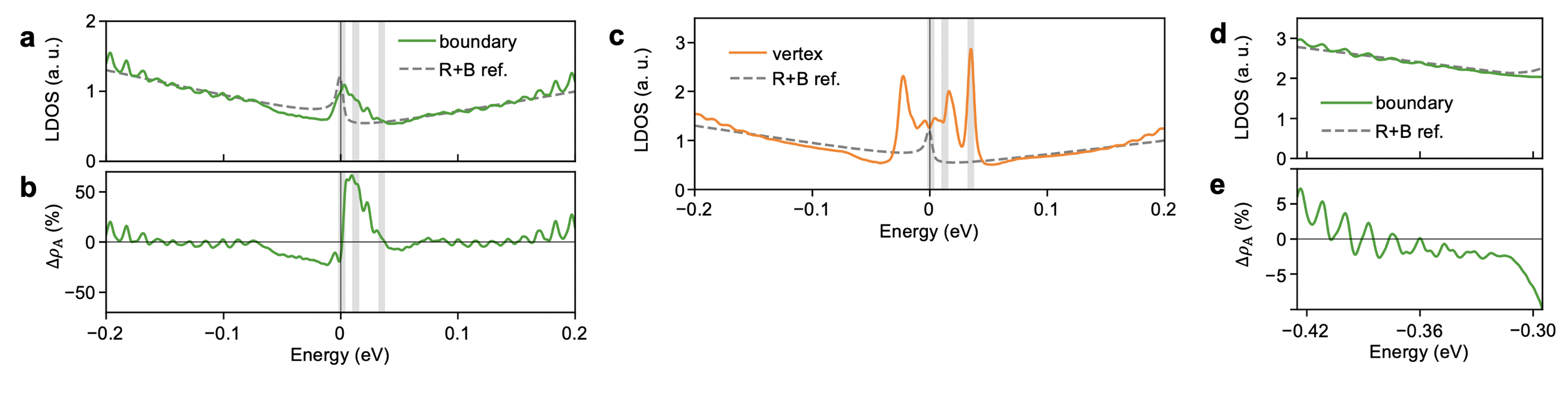}
\caption{
\textbf{LDOS of the TMB system projected on boundaries and vertex.}
\textbf{a} The LDOS ($\rho_A(E)$) projected on the boundary regions of the TMB system (rectangles in Fig.~\ref{fig1}\textbf{c}) is shown by a solid line, and the R+B reference DOS is shown by a gray dashed line. Similarly to the Fig.~\ref{fig2}, the corresponding normalized deviation $\Delta\rho_{A}$ is shown in \textbf{b}.
\textbf{c} The LDOS of the vertex shown with the R+B reference DOS. 
\textbf{d} and \textbf{e} The same data as \textbf{a} and \textbf{b}, but for different energy range. 
The gray shades are $\pm\sigma$ range around the energy levels corresponding to Extended Data Fig.~\ref{EFig_rho_TMB}.
}
\label{EFig_LDOS_TMB}
\end{figure*}

\begin{figure*}[h!]
\centering
\includegraphics[width=1.0\textwidth]{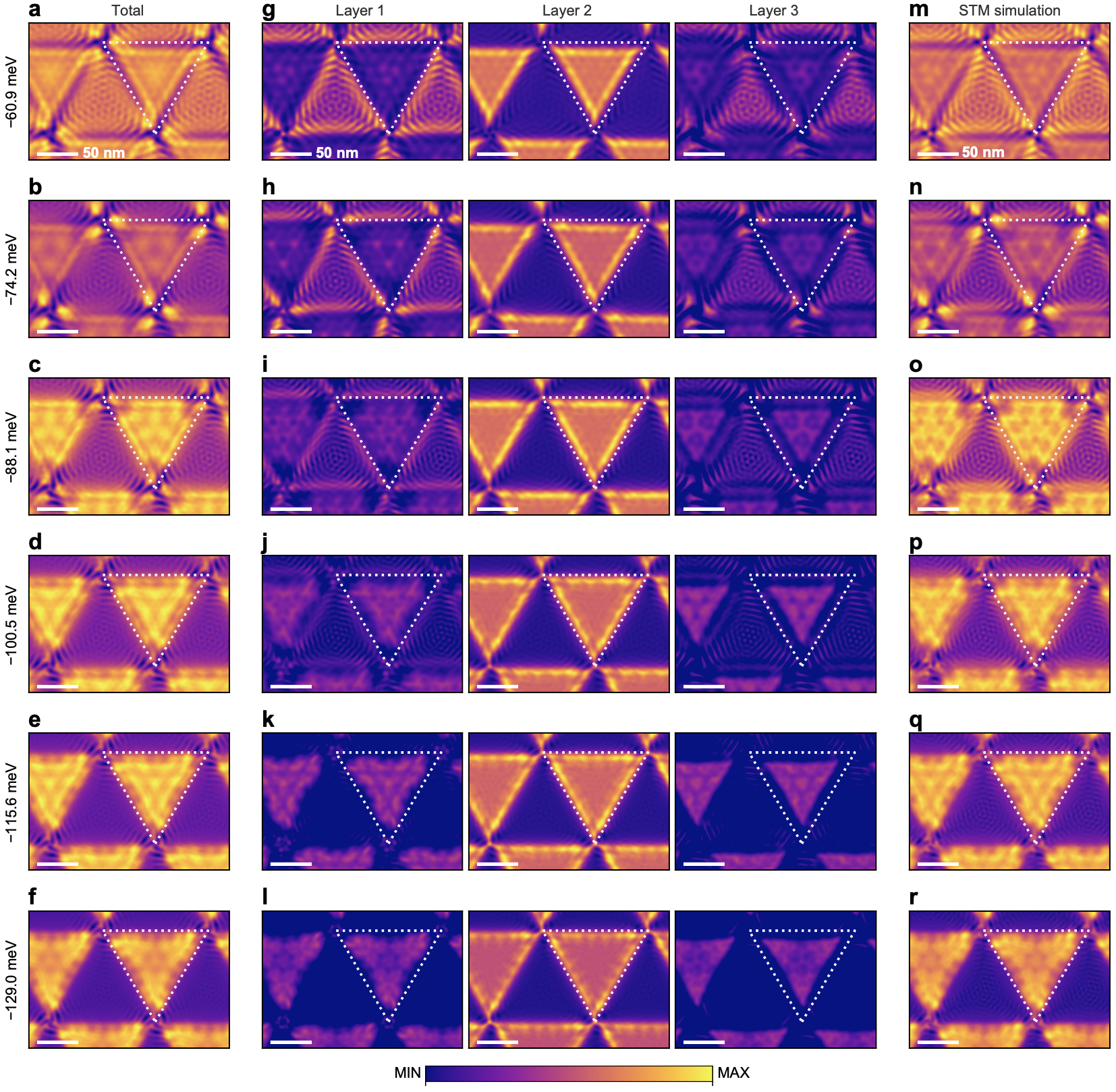}
\caption{
\textbf{Layer- and energy-resolved electron density distributions of the TMB system.} 
\textbf{a}--\textbf{f} Energy-resolved total electron density distributions including contributions from all three layers,
$\rho_E^T({\bf r})=\rho_E^1({\bf r})+\rho_E^2({\bf r})+\rho_E^3({\bf r})$ 
where $\rho_E^i({\bf r})$ denotes the energy-resolved density distribution of layer $i$ ($i=1,2,3$). These energies correspond to the first six panels for $\rho_E({\bf r})=\rho_E^1({\bf r})+\rho_E^3({\bf r})$ in Fig.~\ref{fig3}. All panels within a given row are evaluated at the same energy $E$, indicated on the leftmost side of the row. The colormap in each panel is independently normalized to its own minimum and maximum density values.
Each panel's MIN/MAX value is: 
\textbf{a} 0.629 , \textbf{b} 0.606 , \textbf{c} 0.694 , 
\textbf{d} 0.676 , \textbf{e} 0.673 , \textbf{f} 0.666.
\textbf{g}--\textbf{l} Layer-resolved decomposition of the total energy-resolved density distributions shown in \textbf{a}--\textbf{f}. Each panel contains the contributions from the three individual layers. Within a given panel, all three subpanels share a common colormap range determined by the global minimum and maximum among them, whereas different panels are normalized independently.
Each panel's MIN/MAX value is: 
\textbf{g} 0.475 , \textbf{h} 0.489 , \textbf{i} 0.555 , 
\textbf{j} 0.565 , \textbf{k} 0.594 , and \textbf{l} 0.586.
\textbf{m}-\textbf{r} 
Simulated scanning tunneling microscopy (STM) topographies obtained from the layer-resolved densities following Ref.~\cite{joucken2015carbon}. We employ the constant-current mode with decay length $\lambda=5$~\AA\ and interlayer spacing $d=3.3724$~\AA, assuming an STM current at height $z$ of
$I({\bf r},z)\propto\sum_{i=1}^3\rho^i_E({\bf r})e^{-(z+(i-1)d)/\lambda}$. 
We consider the case where the twisted monolayer (layer 1) is the topmost layer.
}
\label{EFig_rho_layer}
\end{figure*}

\begin{figure*}[h!]
\centering
\includegraphics[width=0.50\textwidth]{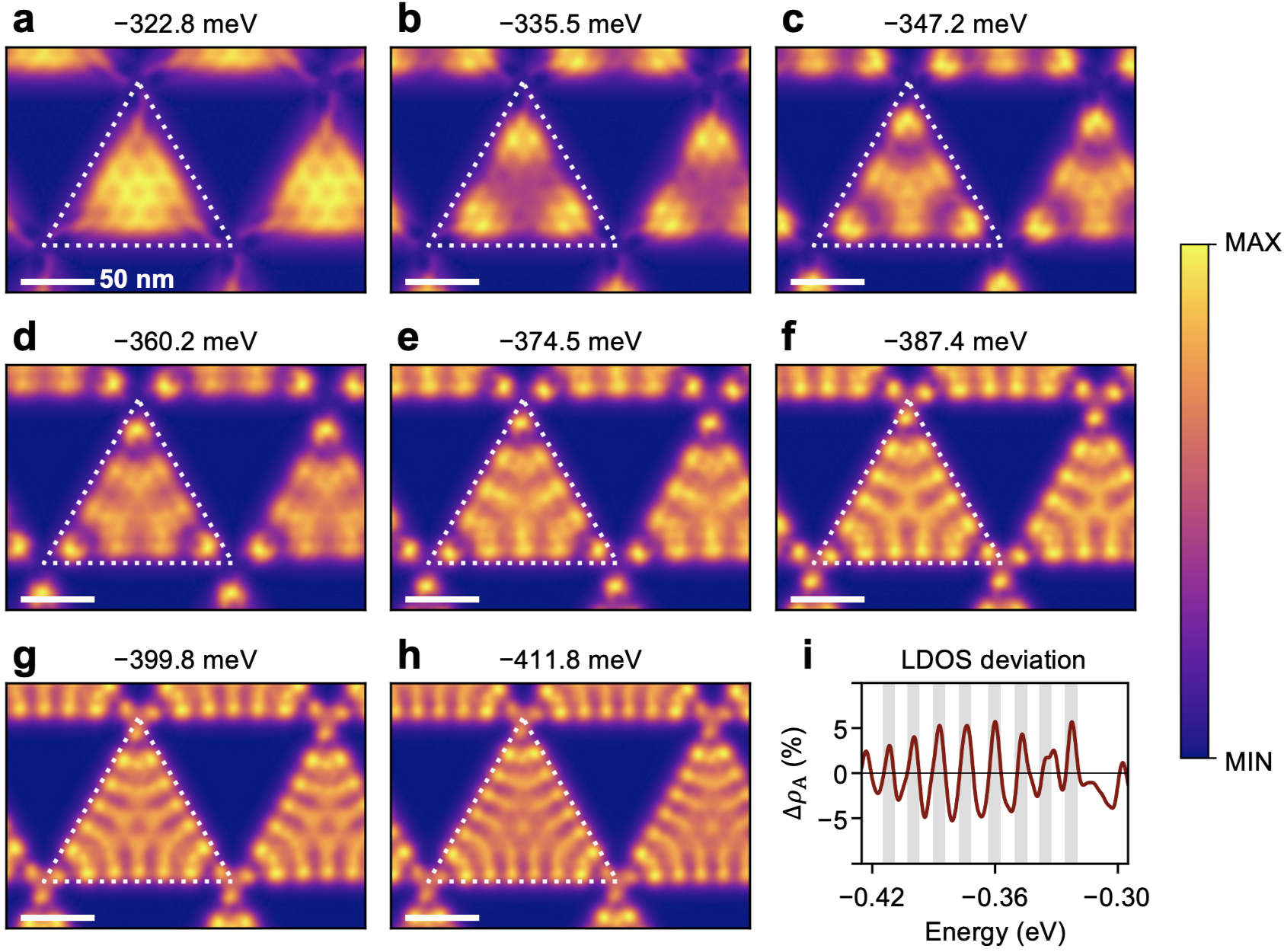}
\caption{
\textbf{Energy-resolved electron density distributions $\rho^T_{E}(\mathbf{r})$ of the TMB system in range 2.} 
The energy levels $E$ above panels \textbf{a}-\textbf{h} are where the density $\rho^T_{E}(\mathbf{r})$ is evaluated for and $E\pm 3$ meV ranges are shown by shades in \textbf{i}, which is identical data as Fig.~\ref{fig2}{\bf f}, the rhombohedral domain LDOS deviation. 
For the $\rho^T_{E}(\mathbf{r})$ in range 2, the contributions from all three layers are included.
The dotted line triangle indicates the rhombohedral domain.
The maximum and minimum of the colormap correspond to each panel's maximum and minimum value of density. 
Each panel's MIN/MAX value is: 
\textbf{a} 0.869 , \textbf{b} 0.853 , \textbf{c} 0.833 , 
\textbf{d} 0.817 , \textbf{e} 0.823 , \textbf{f} 0.821 , 
\textbf{g} 0.816 , and \textbf{h} 0.815.
 }
\label{EFig_ran2}
\end{figure*}

\begin{figure*}[h!]
\centering
\includegraphics[width=0.50\textwidth]{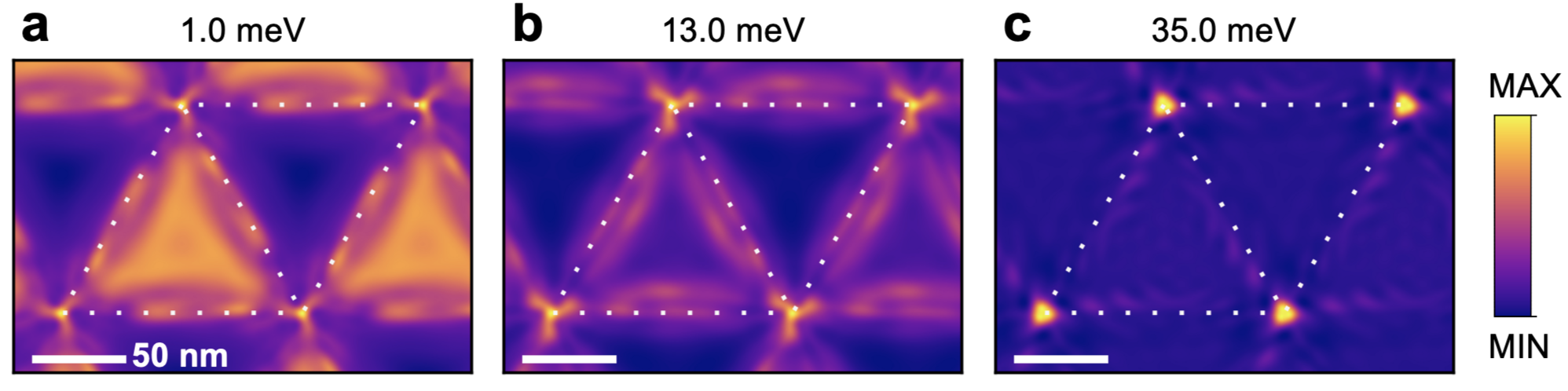}
\caption{
\textbf{Energy-resolved electron density distributions $\rho_{E}(\mathbf{r})$ of the TMB system.} 
The energy level $E$ above each panel is where the density $\rho_{E}(\mathbf{r})$ is evaluated. Those $E$ values are also indicated in Extended Data Fig.~\ref{EFig_LDOS_TMB} \textbf{a-c}.
In this figure, the contributions from every layer are included.
The sparse dotted line indicates the domain boundaries.
Each panel's MIN/MAX value is: \textbf{a} 0.256, \textbf{b} 0.213, and \textbf{c} 0.161.
}
\label{EFig_rho_TMB}
\end{figure*}

\begin{figure*}[h!]
\centering
\includegraphics[width=0.50\textwidth]{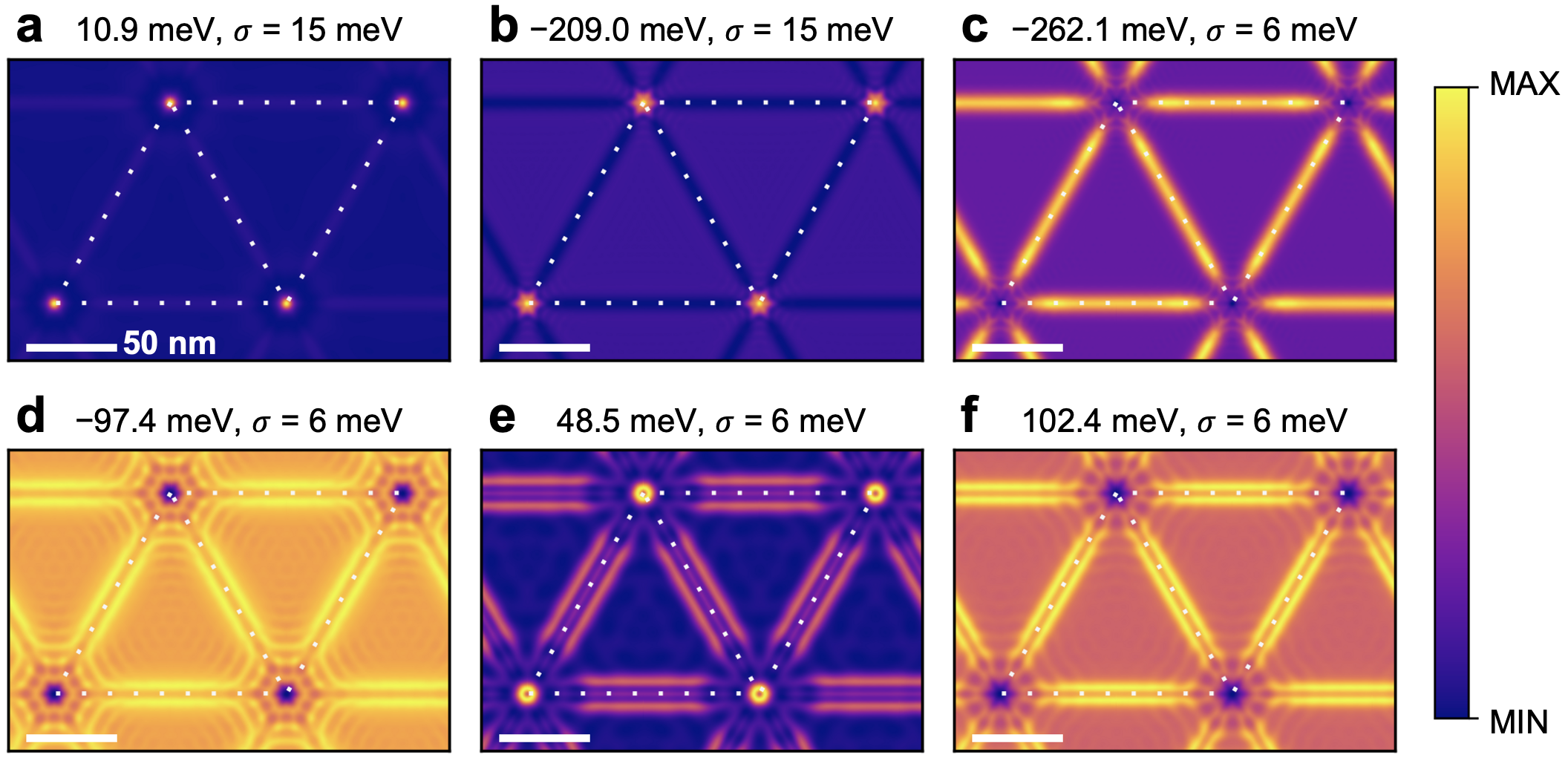}
\caption{
\textbf{Energy-resolved electron density distributions $\rho_{E}(\mathbf{r})$ of the ATT system.} 
The energy level $E$ above each panel is where the density $\rho_{E}(\mathbf{r})$ is evaluated, and the following $\sigma$ determines the width of the Gaussian weight for the integral. Those $E$ values are also indicated in Extended Data Fig.~\ref{EFig_LDOS_ATT} \textbf{c,d} for the boundary-localized densities \textbf{c-f} of this figure, and in Extended Data Fig.~\ref{EFig_LDOS_ATT} \textbf{e} for the vertex-localized densities \textbf{a,b} of this figure.
Note that among the boundary localized densities, \textbf{c} shows a single-wall shape, \textbf{d,f} a double-wall shape, which resembles a boundary state reported for bilayer graphene~\cite{yin2016natcomm}, and \textbf{e} a triple-wall shape.
In this figure, the contributions from every layer are included.
The sparse dotted line indicates the domain boundaries.
Each panel's MIN/MAX value is: \textbf{a} 0.062, \textbf{b} 0.475, \textbf{c} 0.561, \textbf{d} 0.278, \textbf{e} 0.388, and \textbf{f} 0.245.
}
\label{EFig_rho_ATT}
\end{figure*}

\begin{figure*}[h!]
\centering
\includegraphics[width=1.0\textwidth]{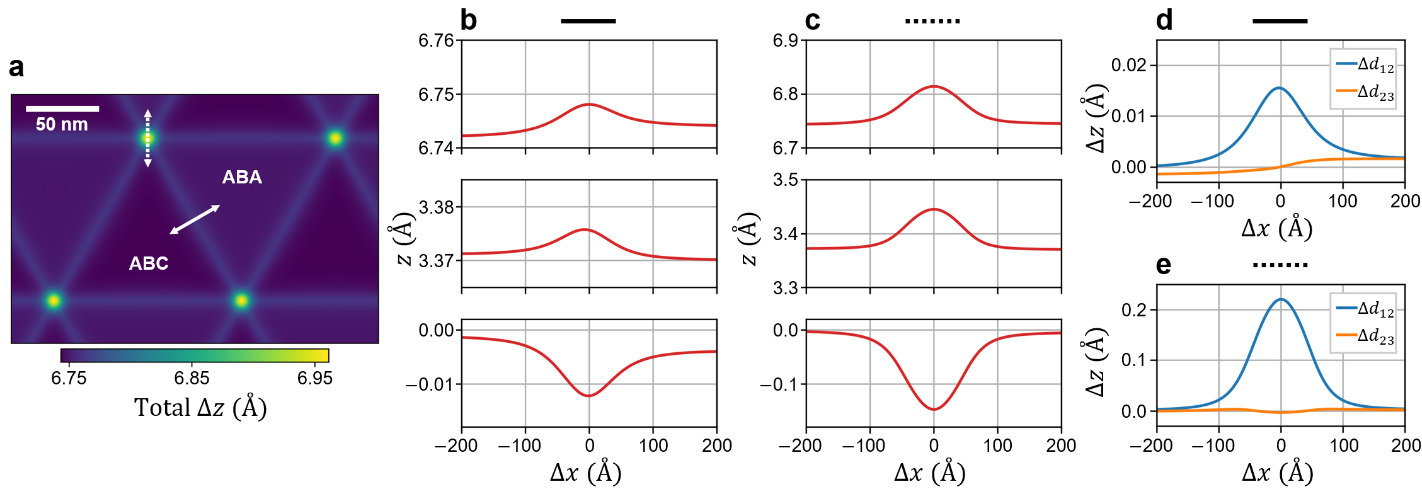}
\caption{
\textbf{Height corrugations across the boundaries.}
\textbf{a}, The spatial variation of total thickness of the TMB for the same area as Fig.~\ref{fig1}{\bf c}.
A solid (dotted) arrow indicate the line along which height variations of each layer are plotted. 
{\bf b} and {\bf c}, From the bottom to top panels, the height variations ($d_i$, $i=1,2,3$) across the boundaries along the solid and dotted arrows in {\bf a}.  
{\bf d}, Along the solid arrow in {\bf a}, the variation of interlayer distance between the bottom and middle ($\Delta d_{12}=d_1-d_2-d^0_{int}$)
and one between the middle and top layer ($\Delta d_{23}=d_2-d_3-d^0_{int}$) with respect to those ($d_{int}$) at the center of domain. 
{\bf e}, The same as {\bf d}, along the dotted arrow in {\bf a}.
}
\label{EFig_height}
\end{figure*}

\begin{figure*}[h!]
\centering
\includegraphics[width=0.85\textwidth]{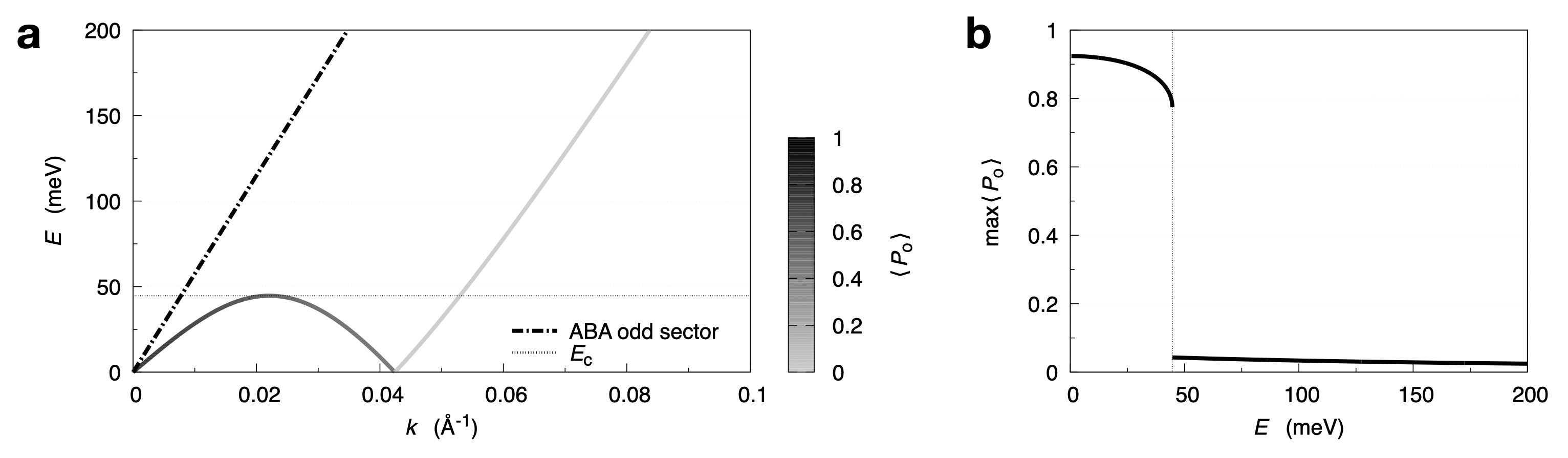}
\caption{
\textbf{Energy bands of ABB stacking and their overlap with massless states in ABA stacking.}  
\textbf{a},
Energy bands along $k_x$ direction for odd sector (massless bands) in ABA stacking (dash-dot line) and ABB stacking (solid lines), respectively. For ABB bands, the grey scale denotes the weight of projection operator $\langle P_o\rangle \equiv \langle \psi_{n{\bf k}}^{ABB}|P_0|\psi^{ABB}_{n{\bf k}}\rangle$ of ABA stacking, $P_o=|A_-\rangle\langle A_-|+|B_-\rangle\langle B_-|$
where $|A_-\rangle =\frac{|A_1\rangle -|A_3\rangle}{\sqrt{2}}$
and $|B_-\rangle =\frac{|B_1\rangle -|B_3\rangle}{\sqrt{2}}$. 
Here $|A_{1(3)}\rangle$ and $|B_{1(3)}\rangle$ are eigenstates of $A$ and $B$ sublattices of first (third) layer. The horizontal dotted line denotes the local maximum energy of $E_c=44.8$ meV at $k=0.022\AA^{-1}$. 
{\bf b}, For a given energy of $E$, the maximum value of projection among $\langle P_o\rangle$ is plotted. The vertical dotted line is $E_c$ in {\bf a}. 
}
\label{Efig_ABB}
\end{figure*}

\begin{figure}[b]
\centering
\includegraphics[width=0.85\textwidth]{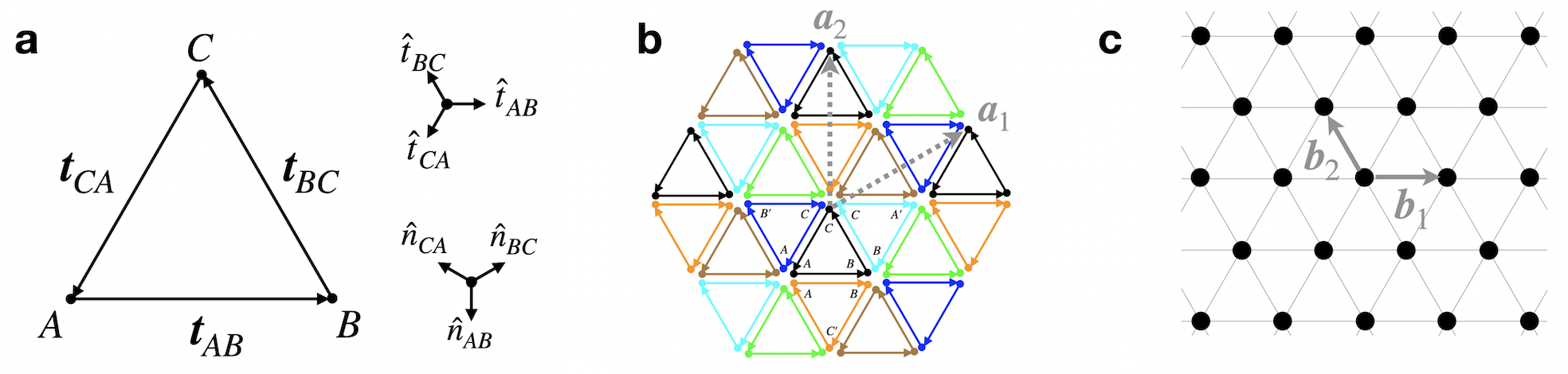}
\caption{
\textbf{Schematic geometries of triangular quantizations.}
\textbf{a}, An equilateral triangle $ABC$ of which vertices are located on 
$A=(0,0)$, $B=(L_M,0)$ and $C=(L_M/2,L_M\sqrt{3}/2)$.
For each side, 
$|{\bf t}_{AB}|=|{\bf t}_{BC}|=|{\bf t}_{CA}|=L_M$
and $L_M$ is a moir\'e periodicity.
$\hat{t}_S$ where $S=AB,BC,CA$ is a unit vector for each side and $\hat{n}_S$ is its normal vector. 
{\bf b}, A triangular lattice constructed by mirror reflections of $\triangle ABC$ with respect to its sides, e.g., $\triangle ABC$ (black color) transforms to $\triangle AB'C$ (blue) through a mirror reflection with respect to side $CA$. Each color denote the same triangle. ${\bf a}_1$ and ${\bf a}_2$ are unit vectors for the lattice. 
{\bf c}, A reciprocal lattice of the triangular tiling in ${\bf c}$ and its unit vectors, ${\bf b}_1$ and ${\bf b}_2$. 
}
\label{EFig_triangle}
\end{figure}

\end{document}